\documentclass[12pt,preprint]{emulateapj}
\usepackage{amsmath}
\usepackage{graphicx}
\usepackage{natbib}
\usepackage{amssymb}
\usepackage{epstopdf}

\usepackage[T1]{fontenc}
\usepackage{amsmath}
\usepackage{oldlfont}
\usepackage[latin1]{inputenc}
\usepackage[english]{babel}
\usepackage{longtable}
\usepackage{rotating}
\usepackage{makeidx}
\makeindex \hfuzz2pt
\usepackage{rotating}
\usepackage{latexsym}
\usepackage{amsthm}
\usepackage{eucal}

\newcommand{\mathsym}[1]{{}}

\shorttitle{Beaming neutrino along the Earth}
\shortauthors{Fargion $\&$ D'Armiento \textit{et al.}}

\begin{document}


\title{Beaming neutrino and antineutrinos across the Earth  \\
to disentangle neutrino mixing parameters\\
}


\author{Daniele Fargion\altaffilmark{1,}\altaffilmark{2} and Daniele D'Armiento\altaffilmark{1}\\
 Paolo Desiati\altaffilmark{3}, Paolo Paggi\altaffilmark{1}}

\altaffiltext{1}{Physics Department, Rome University 1, P.le A Moro 5,Rome Italy; daniele.fargion@roma1.infn.it}
\altaffiltext{2}{INFN Rome Univ. 1, Italy}
\altaffiltext{3}{Department of Physics, University of Wisconsin - Madison, 1150 University Avenue Madison, WI 53706, U.S.A.; paolo.desiati@icecube.wisc.edu}

\begin{abstract}
A MINOS result  \cite{1} seemed to hint  a  different anti-neutrino mass splitting and  mixing angle with respect to the neutrino ones, suggesting a CPT violation in lepton sector.  However more recent MINOS data    \cite{111} reduced the $\nu_{\mu}$-$\bar{\nu}_{\mu}$ differences, leading to a narrow discrepancy nearly compatible with no CPT violation.  However last  years of OPERA activity on tau appearance is still un-probed (one unique event) and a list of parameters ( $\mu$-$\tau$ flavor mixing, tau appearance, any eventual CPT violation, $\theta_{13}$ angle value and any hierarchy neutrino mass) mAy  need more tools to be disentangled. Atmospheric anisotropy in muon neutrino spectra in Deep Core, at ten-tens GeV (yet unpublished), can hardly reveal eventual $\nu_{\mu}$-$\bar{\nu}_{\mu}$  oscillation parameters asymmetry. We considered here how the longest baseline neutrino oscillation available, crossing most of the Earth diameter, may improve the measurement and disentangle at best any hypothetical CPT violation within earliest (2010) and present (2012) MINOS bounds (with 6$\sigma$ a year), while testing at highest rate $\tau$ and even the $\bar{\tau}$ appearance.  The  $\nu_{\mu}$ and ${\bar\nu_{\mu}}$  disappearance correlated with tau appearance is  considered for those largest distances. We thus propose  a beam of $\nu_{\mu}$ and ${\bar\nu_{\mu}}$ crossing through the Earth, within an OPERA-like experiment from CERN (or FermiLAB), in the direction of ICECUBE-DeepCore $\nu$ detector at the South Pole. The ideal energy lay at 21 GeV energy, to test the disappearance or (for any tiny CPT violation)  the partial $\bar\nu_{\mu}$ appearance.  Such a tuned detection experiment may lead to a strong signature of  $\tau$ or $\bar{\tau}$ generation even within its neutral current noise background  events: nearly one $\bar{\tau}$  or two $\tau$ a day. The tau appearance signal is above (or within) 10$\sigma$ a year, even for $1\%$ OPERA-like experiment. Peculiar configurations for $\theta_{13}$ and hierarchy neutrino mass test may also be better addressed by a Deep Core-PINGU array detector   beaming  $\nu_{\mu}$ and observing $\nu_{e}$ at 6 GeV neutrino energy windows.
\end{abstract}



\section{Introduction: is there an eventual neutrino CPT violation?}

In the middle of the last century unexpected broken symmetries in elementary interactions have been discovered:
 first  the violation of Parity-symmetry was observed (events as seen as on a mirror are not always occurring in Nature), as well as a Charge-symmetry violation in phenomena that involve the weak force. In the $70$'s Parity-Charge (PC) violation was also observed in elementary interactions; this result, assuming a CPT invariance, implied a Time-asymmetry as well. Up to now CPT invariance has been seemingly solid and able to survive most tests and the basic theoretical needs. However, the recent MINOS observations \cite{1} seemed to imply (or now at least to marginally hint \cite{111}) a different anti-neutrino mass splitting with respect to well known neutrino one, leading to a possible CPT violation. Moreover the measure of $\nu_{\tau}$ and ${\bar\nu_{\tau}}$ may also test, once revealed by a meaningful statistical rate, such eventual broken symmetry.
 There are various severe constraints on CPT violation from the neutral Kaon oscillation (an a-dimensional mass discrepancy  at $8\
  10^{-18}$ level), as well as, at lower level from the charged lepton sector (at $9\cdot 10^{-9}$ level for electron pair masses) \cite{3}. Therefore a significant (or even marginal) neutrino anti-neutrino mass difference may open new roads in our particle physics understanding \cite{4}. This CPT violation might indicate a very peculiar role of neutral leptons  in matter/anti-matter genesis, and it may address unsolved lepton-baryon-genesis open puzzle, related to cosmological mysteries. Consequently  such a CPT violation, if confirmed, might become one or the main (amazing) discovery of the century. Therefore, even though last MINOS observation may be  consistent with CPT conservation,
we nevertheless have  the duty to carefully  inquire if such CPT violation may be observed by detecting atmospheric neutrinos additional anomaly or by proposing a new experiment aimed to disentangle this possibility.

 The main phenomenon of our proposal is to reveal discrepancy among $\nu_{\mu} \rightarrow \nu_{\tau}$, $\nu_{\mu} \rightarrow \nu_{\mu}$  and ${\bar\nu_{\mu}} \rightarrow {\bar\nu_{\tau}}$, ${\bar\nu_{\mu}} \rightarrow {\bar\nu_{\mu}}$  that it is the flavor mixing  both in its tau appearance and in its muon  disappearance at the Deep Core detector. An introduction on expected atmospheric neutrino spectra in DeepCore is our first step into this field. Our recent and present foreseen Deep Core records are not identical to their ones (our are nearly less than half and our peak energy is at half channel number). But more than one year neutrino records by DeepCore, soon-to-be published, may shed light in this theoretical estimates.

  Let us remind that the tau lepton birth need a high (4-5 GeV) energy birth threshold; the best muon-tau neutrino conversion available at Earth is around twenty GeV because the neutrino oscillation is bounded by the finite Earth size leading us to the present (somehow constrained) longest baseline proposal centered at $21$ GeV. We did consider the largest existing facilities (emission: CERN, Fermi Lab; detection: SK, DeepCore) at a minimal costs.
  Only a new bent neutrino tunnel (let say at $1\%$ length-flux regime respect to OPERA) will be needed. The estimate of the beaming facilities and the estimate of the expected neutrino flux-signals is our main goal,  whose statistical significance is finally discussed. The main signature of the $\tau$ (or $\bar{\tau}$) lepton appearance will not be a well defined tau track as it is in OPERA emulsion, but a statistical shower rate enhanced by taus respect neutral current  (or electrons) events.

\subsection{A brief historical remind}
  Neutrino oscillations were first thought nearly half a century ago by \cite{-2} (in prompt resonance with an analogous proposal by Cabibbo for the hadron sector \cite{-6}). The $\nu_{\tau}$ and $\bar{\nu_{\tau}}$ are the most recently discovered neutral leptons (ten years ago \cite{-5}), correlated to the heaviest charge lepton, the $\tau$, discovered only 37 years ago \cite{-1}. Their eventual  CPT violated masses would be somehow surprising. Indeed, such a  detection would require extraordinary evidence. For this reason the Longest Baseline under consideration offers a sharp test (possibly the best one up-to-date)  to disentangle hidden tiny parameters even  within last  mild MINOS CPT results \cite{111}.

\section{Main idea to disentangle mixing flavors }
We considered first an ongoing experiment based on atmospheric neutrino signal in DeepCore that may be somehow a bench-mark of the MINOS CPT detection; however the muon track energy measure and the energy-angular resolution might confuse the (mild) expected CPT anisotropy \cite{111} in Deep Core map and spectra. Therefore we  focus here on a future possible ad-hoc Long Baseline experiment able to sharply  confirm the CPT violation in a very short time and accurate way. We studied the appearance of any anti-neutrino-generated muon in the energy-distance range where CPT conserved oscillation is almost vanishing, while CPT violated oscillation is (partially) allowed based on the experimental parameters determined by MINOS. By doing this we will be able also to test with great accuracy the muon-tau mixing, leading also to a very precise estimate of their mixing parameters that may shed light to a possibly hidden symmetry unwritten into a  tuned value:  $\sin(2 \theta_{23})\simeq1$.

\begin{figure*}
\begin{center}
\includegraphics[angle=0,scale=.15]{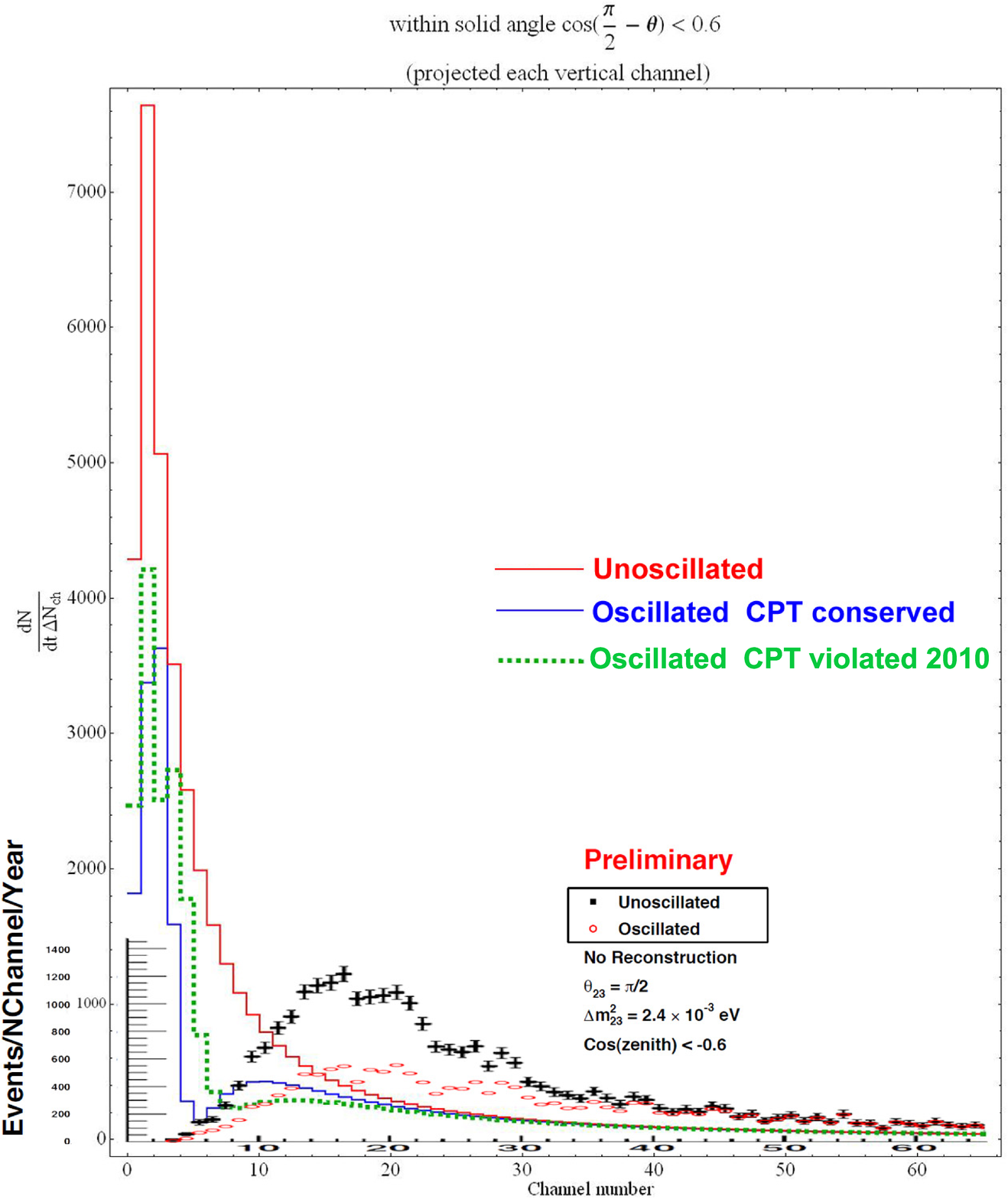}
\includegraphics[angle=0,scale=.15]{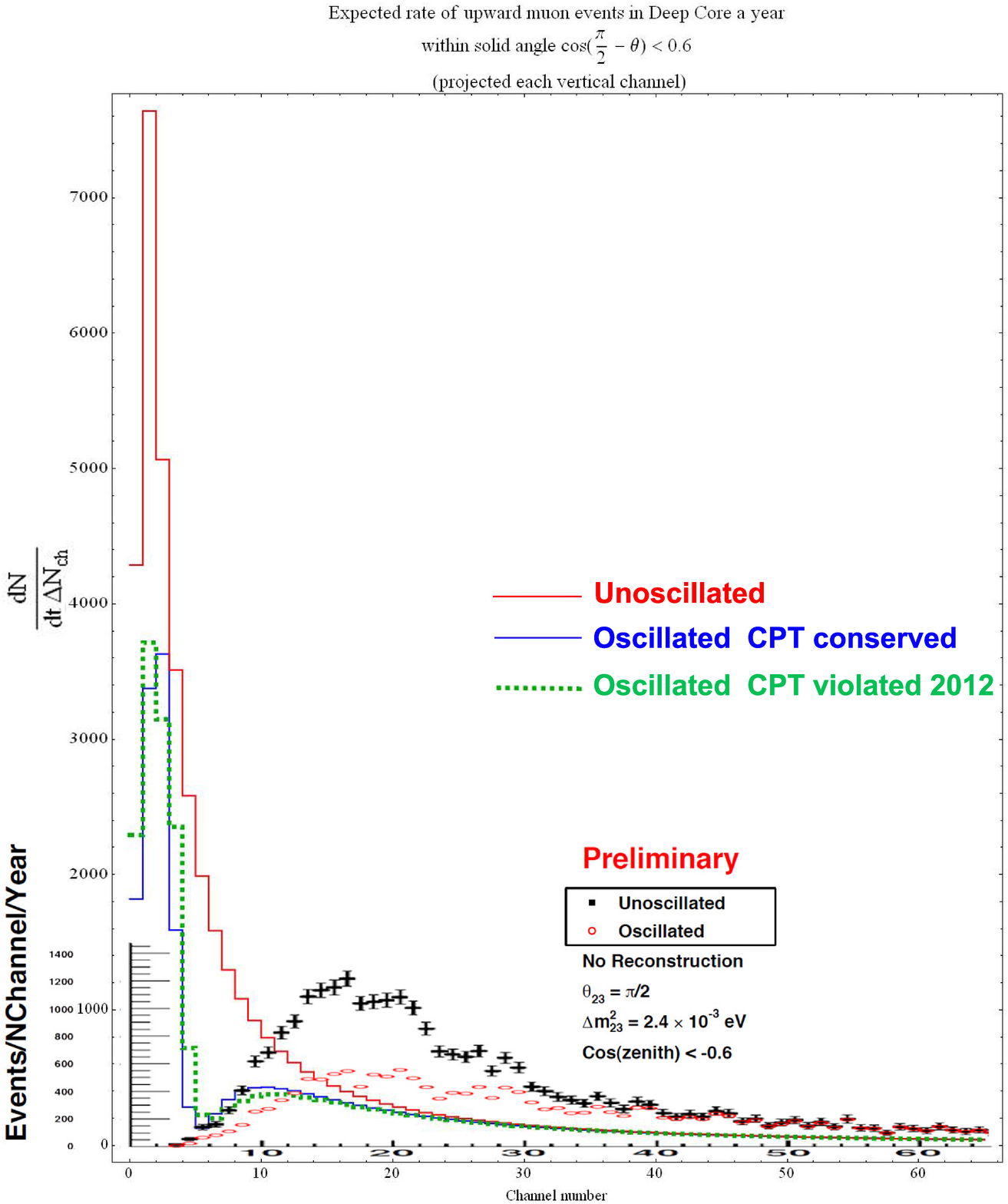}
\caption{ Our expected atmospheric neutrino spectra in Deep Core, as a function of the DOM channel number, both for earliest (2010) and late (2012) CPT parameters. Here we apply the Earth matter influence along the overall oscillation, versus the DeepCore preliminary expectations (2009-2010). Note (left side) the remarkable deviation (our green dotted curve versus DeepCore small red ovals) in 2010 MINOS CPT scenario (respect CPT conserved rate in  blue-thin line) and the present (unique dotted green deviation at channel n.4) in last right handed figure. While CPT MINOS deviation was on 2010 -in principle- well detectable because a suppression (respect CPT conserved case) by $30\%$ around channel 8-12, the deviation in small CPT deviation on 2012 is unobservable.  There are great differences among our spectra and DeepCore expected ones in the low energy ranges;  a mild difference at high energy. The main feature of our spectra  (based on Super Kamiokande scaled data) and DeepCore one is  the low rate at high energy (our half of the DeepCore ones), the flux maxima (for us at channel 10 and for DeepCore  at channel 20) and the remarkable higher flux at a few channels (3-5).    The DeepCore simulation  \cite{0} has been used and shown here as a reference bench mark. Their data are still unpublished. Our predictions are very clear and testable in a near future.}\label{01}
\end{center}
\end{figure*}

In the present  article  we did not mention the vacuum (more known) mixing, but we discuss the mixing within the Earth keeping care of the matter density. At $20 - 50$ GeV, nevertheless,  the flavor mixing along the Earth diameter is  not very much different from the vacuum case; the neutrino signals  may be recorded in DeepCore array and its volume exhibits an effective detection mass of ($4-15$ Mton) in that energy range (see left side Fig. \ref{20}). The starting reason to consider inner IceCube DeepCore as the candidate  detector for muon-tau mixing is its huge volume (4000 times OPERA) and its complete oscillation distance from Cern.

\begin{figure}[!h]
\begin{center}
\includegraphics[angle=0,scale=.3]{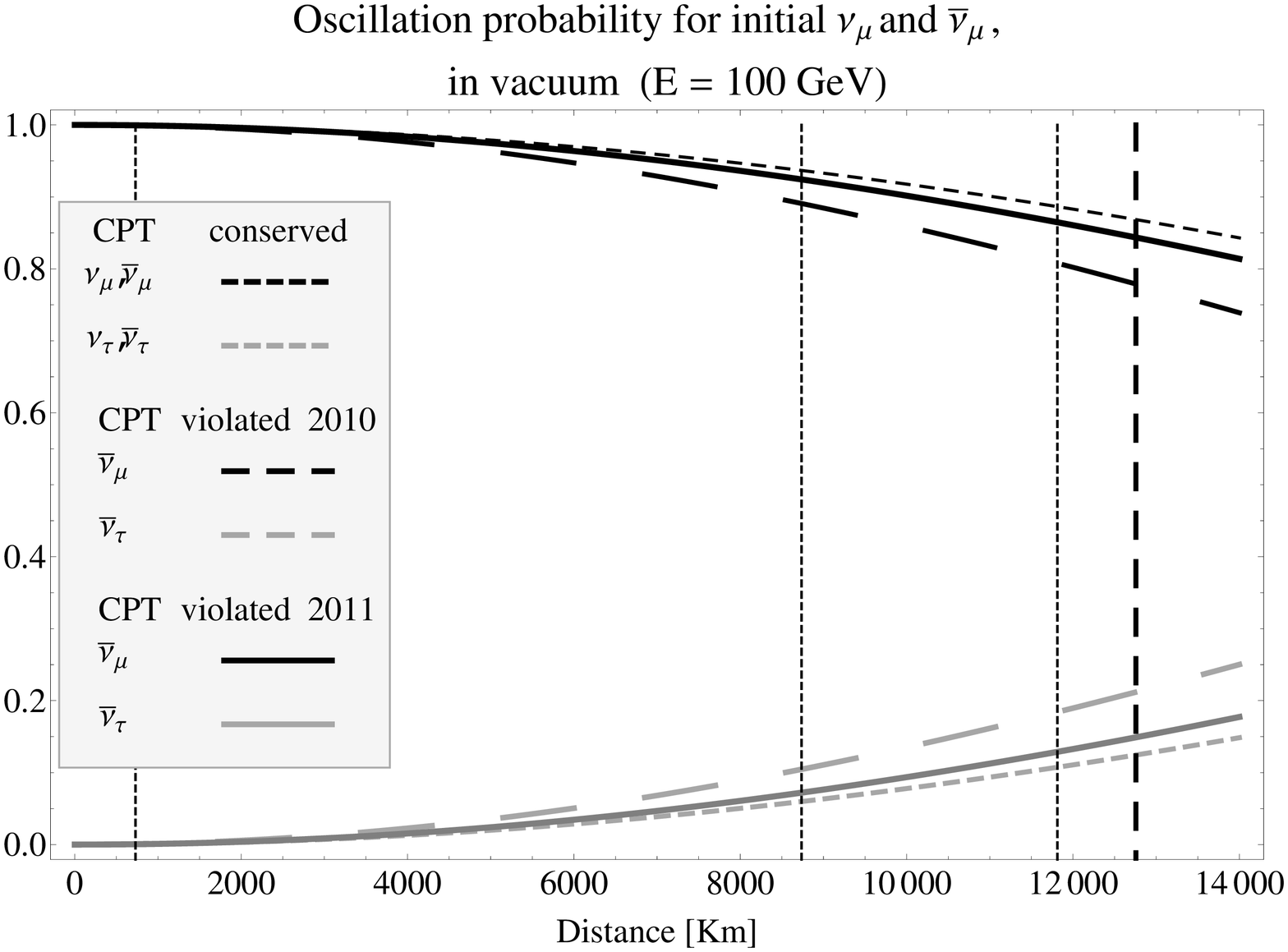}
\includegraphics[angle=0,scale=.3]{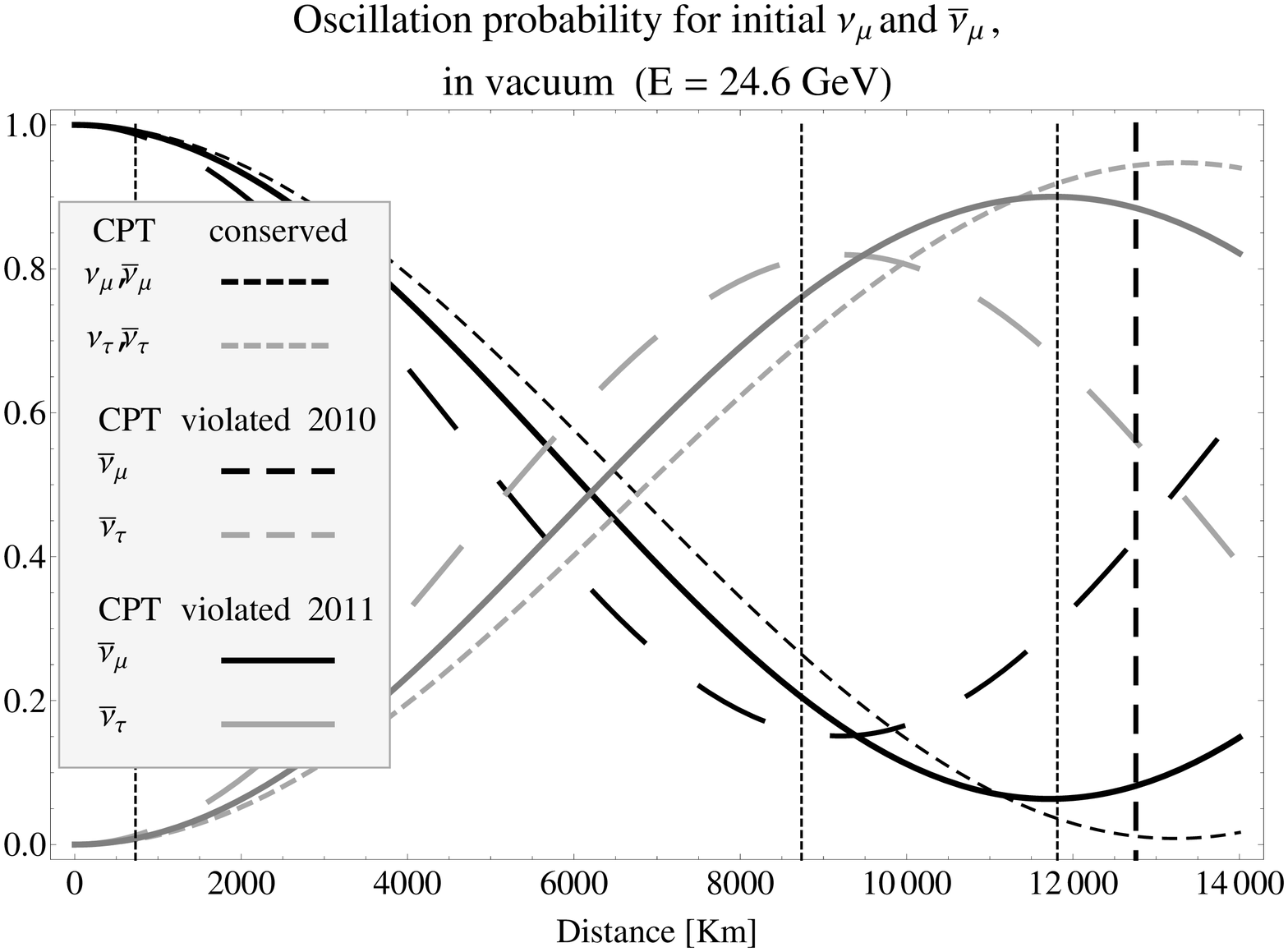}
\caption{Neutrino and anti-neutrino Muon  oscillation probability in ($\nu_{\mu}\rightarrow\nu_{\tau} $) \emph{and the anti-neutrino} muon ($\overline{\nu}_{\mu}\rightarrow\overline{\nu}_{\tau} $) oscillation probability for three cases: a first conserving CPT symmetry,  $\sin(2\theta_{23})=\sin(2\overline{\theta}_{23})=1$ and  $\Delta m_{23}^{2}=\Delta \overline{m}_{23}^{2}=(2.35^{+0.11}_{-0.08})\cdot 10^{-3}eV$ (CPT conserved case), a second one violating CPT symmetry \cite{1}, where its anti-muon neutrino CPT parameters are: $\Delta \overline{m}_{23}^{2} =(3.36^{+0.45}_{-0.40}($stat.$)\,\pm\,0.06($syst.$))\cdot 10^{-3}$ eV$^{2}$ and $\sin^{2}(2\overline{\theta}_{23})\,=\,0.86\,\pm\,0.11($stat.$)\,\pm\,0.01($syst.$)$, and a third (last) CPT symmetry \cite{111} where: $\Delta \overline{m}_{23}^{2} = 2.62 \cdot 10^{-3}eV$ and $\sin^{2}(2\overline{\theta}_{23})\,=\,0.945$.
These values are applied along all the article. These cases are shown only here in the simplest vacuum approximation (whose results at high energy are nearly coincident with the matter ones to be applied later). On the top, the  neutrino mixing  at 100 GeV, in vacuum, at IceCube detection threshold, is quite negligible. A tiny atmospheric up-going  muon neutrino suppression will be reflected in a tiny anisotropy hardly to be detected because ICECUBE events are ruled mostly by TeV ones. At bottom figure the lowest $24.6$ GeV case shows  a clear variability and anisotropy due to the muon disappearance to be  observed in DeepCore.  An average anisotropy at tens GeV maybe detected as in next figures.  The vertical dotted lines refer to the baseline distances for OPERA, SuperK and ICECUBE, or across all the Earth as it is in Table 1.}\label{02}
\end{center}
\end{figure}

Let us summarize the main questions and the aims of this paper and its structure. The recent MINOS bounds on CPT violated parameter  may be slightly improved by  MINOS-like experiments, such as OPERA at Gran Sasso, for instance. The limitations are related to the used low energies  (a few GeV) in order to observe oscillation within the baseline distance of about $735$ km of the FNAL (Fermi National Accelerator Laboratory)-MINOS or the CERN-OPERA experiments. The higher the energy, the larger the distance needed to observe a complete oscillation, but also the better the neutrino beaming, because of the high Lorentz factor of pion decay, as well as the larger the neutrino cross-section.
Incidentally the approximate beaming solid angle shrinks by a factor proportional to  $E_{\pi}^{2}$, and the neutrino-matter cross-section grows as $E_{\nu}$ providing a global signal enhancement amplified by a factor proportional to $\sim E_{\nu}^{3}$.
Therefore a long-baseline experiment at 22 GeV, approximately the present  energy threshold of CERN-DeepCore in IceCube discussed below, may play a better role ($8000$ times better than 1 GeV experiment and a factor 2 because of better energy beaming respect $17$ GeV energy) to define oscillation parameters. Moreover the dilution factor due to the much greater distance from CERN of DeepCore than OPERA, a factor $\simeq 240$, is widely compensated by the detector mass ratio (DeepCore  versus OPERA), at least by a factor $\simeq 4800$, implying a benefit of a factor  $\simeq 20$. In addition the larger distance in the longest baseline offers a complete $\nu_{\mu}\leftrightarrow \nu_{\tau}$ conversion with respect to $1.5 \%$ of OPERA, providing a further gain of an additional factor $\simeq 60$.
All together the advantages of a long baseline experiment (see for the beaming deflection maps Fig \ref{Tunnel-CERN}, Fig \ref{Tunnel-Fermi} from CERN and Fermi-Lab sources) with DeepCore (respect to OPERA) in tau appearance,  is a factor of about or above $2400$; finally all the born $\tau$  (within the limited 4.8 Mton DeepCore) will be observable, while the real-to-estimated efficiency ratio happens to be $1:15$ for OPERA, leading to an exceptional ratio $(15\cdot 2400)^{-1} = \frac{1}{36000}$ between OPERA and our test in tau appearance. One tau a day in our scenario at $1\%$ OPERA size versus one tau a year in present OPERA experiment (see more precise details in next Tables). We remind that we are considering half detection volume respect  the one claimed, for prudential reasons \cite{15,12}.

\begin{figure}[htbp]
\begin{center}
\includegraphics[angle=0,scale=.3]{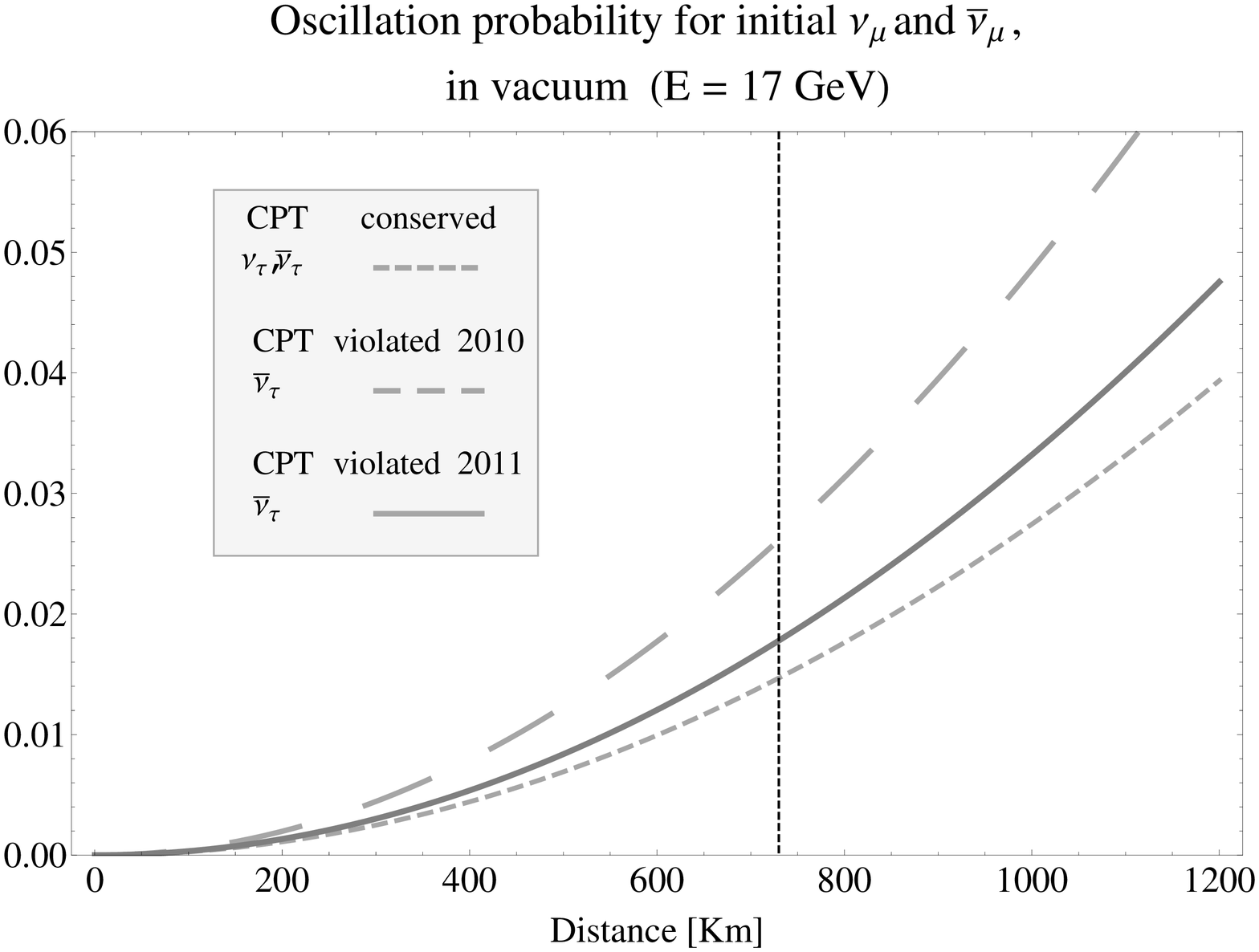}
\caption{At nominal OPERA $17$ GeV energy, neutrino \emph{and the anti-neutrino} muon oscillation probability in ($\nu_{\mu}\rightarrow\nu_{\tau} $)  ($\overline{\nu}_{\mu}\rightarrow\overline{\nu}_{\tau} $)
 oscillation probability, in vacuum, for different zoom distances, both for CPT conserved and CPT violated parameter (\cite{1}and\cite{111}). As mentioned above the mixing, do not differ much from the matter case. Note that the anti-tau appearance is nearly doubling  respect conserved CPT case \cite{1}, but very little increase by twenty percent for last parameters \cite{111}; nevertheless the anti-tau  cross-section depress the final result by a complementary factor (half) leading again  to the same level as one tau appearance a year(\cite{1}), or  just a negligible half a year in \cite{111}. To be statistically more  convincing we considered a beaming from Cern (or FNAL) to IceCube producing a  much larger $\bar{\tau}$ rate, discussed below.}\label{03}
\end{center}
\end{figure}

The idea of long baseline is not totally new.  Previous proposals have been related to smaller distances (at FNAL-Super Kamiokande or CERN-Super Kamiokande) and to the lower neutrino energy aimed to detect a  windows where  the MSW (S. Mikheyev, A. Smirnov, L. Wolfenstein - MSW)
resonance in matter takes place \cite{13}. Here we suggest a different energy window, a different approach and detectors: a) to use the DeepCore detector to test the fluxes of atmospheric neutrino (in CPT conserved versus CPT violated case) at a few tens GeV, b) to  beam  tuned neutrinos in artificial experiment at the largest terrestrial distances (CERN or FNAL toward DeepCore) to disentangle the CPT conserved from the CPT violated case and to focus on mixing flavor values (tau and anti-tau  appearance). The eventual future PINGU experiment whose dense optical module distribution allow a lower energy neutrino threshold maybe important in the lower energy resonance MSW  windows ($\sim 3$, $\sim 6$ GeV),  also of interest for eventual neutrino CP violation parameters as well as in the $\theta_{13}$ detection and inverted neutrino mass hierarchy, see Fig. \ref{Theta-_13}, \ref{IHmu}, \ref{IHe}.

We first consider CERN or FNAL beaming to Super Kamiokande (SK), detecting $\nu_{\mu}$ by their muons mostly produced on a large external mass (partially contained events), due to the so-called muon rock (i.e. those produced outside the water detector). The muon rock signal contribution may significantly increase (mainly for SK) the effective neutrino detection volume and their observable rate, leading to useful results able to disentangle the CPT puzzle only in the earliest CPT MINOS claim \cite{1}, but not in later \cite{111}. In particular we considered here those neutrino energies such that muon (or better anti-muon) neutrinos appear (for CPT violated case) at the same time where (for a CPT conserved physics), they should be almost totally converted and absent (see the bottom  fig. \ref{02} near the muon survival minima). This leads to a clear significant signal-to-noise signature. We concentrated mainly on the expected muon disappearance energy-distance window (and not on the opposite anti-muon CPT violated disappearance)  because the CPT conserved case exhibits a complete mixing with $\sin(2 \theta_{23})\simeq 1$ (the MINOS hypothetical CPT violated parameter \cite{111} lead only to a partial oscillation probability with $\sin(2 \bar{\theta}_{23})\simeq 0.95$,  fig. \ref{02}). A signal over a null background  is more remarkable than any signal over a non vanishing noise.

Secondly and in particular we considered as a neutrino detector the densely instrumented DeepCore apparatus inside the IceCube neutrino
telescope, with an effective volume conservatively estimated to be $1-4.8$ Megaton for $\sim 20$s GeV neutrino energy window (see left of Fig.
\ref{20}, assuming a prudential suppressive factor $2$ at $20$ GeV ). DeepCore at present is the most distant and the largest target to detect
$\simeq 20$s GeV neutrinos. We are aware that this apparatus, at 20 GeV, mostly detects neutrino events with a reconstructed directional
resolution of the order of the degree, and a smeared energy estimation.

 Actually the arrival direction is defined by a wide conical solid angle and its energy is valuable by an approximated windows. However the artificial source may define the neutrino energy, their known distance may offer the correct oscillation factor at reasonable range. Therefore we estimated in detail  for each source (CERN or FNAL), for each proton to target source flux (and its secondary neutrino flux) and for each detector distance (SK or IceCube) the results (muons or tau events) by a chain of  neutrino-signal values source-propagation-flavor mixing and oscillation in Earth, the detection rate in volume inside or outside the detector.  Each value or formula is deeply correlated to the previous one, leading to a realistic estimate of muon or anti-muon signals (as well as tau-anti tau), designed  at best to disentangle  the recent MINOS hypothetical   CPT  asymmetry as well as to fix with high accuracy muon-tau flavor mixing parameters.

We also kept in mind some key practical problems related to the  charged beam bending into a deep underground tunnel, the eventual spectroscopy selection  in a narrow energy band, the external muon (rock-ice) production, the muon and tau detection in deep core string array (see Fig.\ref{SK-DC}, left for SK detector, right for DeepCore Array detector).  All these related evaluations make the paper wide and possibly difficult to embrace. However the main message is simple:  the CPT puzzle (that could be  partially tested  in out-coming DeepCore atmospheric neutrino event rate under former CPT violating parameters),  is not well disentangled by later CPT violating parameters;  in Fig \ref{01}, left side, is shown the atmospheric neutrino spectra for \cite{1} parameters, while in the right side for the new ones \cite{111}. As we shall see anyway the tiny eventual presence of CPT violation maybe disentangled  (above 6 sigma, at 1\% of OPERA size experiment) by our proposal, where neutrino  beaming are tuned  in muon-tau CERN-Deep Core experiment.  Most of the theoretical problem related to the bending and  to the neutrino oscillation in vacuum or in matter have been faced and calibrated (also with other authors' estimates) in present paper.

\section{The structure of the paper}
       In the present section we remind the very recent IceCube atmospheric neutrino results \cite{5}, \cite{6}. We  test our prediction of  DeepCore capabilities to detect atmospheric neutrinos in absence  and, within our  article, in presence of CPT violating terms, based on \cite{7}.  The way to  detect  the muon Cherenkov lights (or anti-muon ones) is related to the photo-tube recording in chain channels.  The DeepCore array, with its vertical parallel strings at  72 m distance  one from another and with 7 m separation between photo-tube sensors along each string, is able to reconstruct more precisely direction and energy of tens GeV events for almost vertical tracks.
 We elaborated our slightly different expectation for cosmic atmospheric signals in DeepCore \cite{-9}, for old \cite{1} parameters and here, See Fig. \ref{01}, also for new ones  \cite{111}. Last muon neutrino MINOS CPT conserved mass splitting in the anti-muon sector is:  $\Delta m_{23}^{2}=\Delta \overline{m}_{23}^{2}=(2.35^{+0.11}_{-0.08})\cdot 10^{-3}$ eV$^{2}$ and $\sin(2\theta_{23})=\sin(2\overline{\theta}_{23})=1$,  versus the larger one for CPT violated case  $\Delta \overline{m}_{23}^{2} =(2.62^{+0.31}_{-0.28}($stat.$)\,\pm\,0.09($syst.$))\cdot 10^{-3}$ eV$^{2}$ and $\sin^{2}(2\overline{\theta}_{23})\,=\,0.945$. The global effect in last CPT violated case for atmospheric neutrino is a tiny enhancement in Deep Core channel 5, and a nearly unchanged behavior near and above Channel $10$.  To see the ideal CPT violated versus conserved neutrino mixing probability, at different energies see Figure \ref{03}.
 These preliminary behaviors had inspired our tuned beaming experiment along present largest labs and detectors at largest distances.
 In following figures \ref{05} we shall consider the largest mixing and tuned energies (for given CERN-SK, CERN-DeepCore distances) able to suppress at a minima the CPT conserved muon neutrino flux. Exactly in those minimal energy windows we will imagine to disentangle any tiny CPT violated anti-neutrino behavior. In the next figures \ref{05}, we review the same tuned oscillations (for SK and DeepCore targets)as function of distance, keeping care of the very subtle role of the matter during neutrino propagation. We emphasize that the low energy (few GeV band) does differ greatly from the vacuum case, while tens GeV neutrino propagation is marginally influenced by the matter presence.  In next figures \ref{06}, on the contrary, we show the oscillation probability as function of the energy, once fixed the given detectors distances (SK and DeepCore) from CERN source; in the latter case the fine tuned MSW resonance play a key role in the few GeV energy band and it is shown but not discussed in detail.
 In the last figure \ref{deltaE10},\ref{deltaE20}, we add to the same probability mixing as a function of the energy a noise due to neutrino beam energy non monochromaticity ($\frac{\Delta E}{E} = 10\%, 20\% $) see right side Fig. \ref{20}. The presence of such energy smearing increases the noise and reduces the signal significance. However as discussed in the corresponding tables, the event rate in worst $1\%$ OPERA-like experiment may lead to a remarkable $6\sigma$ signal detection of an hypothetical CPT violation  within  present  bounds \cite{111}. The correlated parameter map derived by a year of recording (in a minimal $1\%$ OPERA beaming experiment) is somehow (preliminary) shown in the figure \ref{MINOS-ellipse}.


\subsection{Table structures: muon-tau  neutrino detections in vacuum or matter}
The way to estimate the eventual beaming and detection of neutrinos to DeepCore is based on a chain of correlated evaluations that we used and calibrated with known experiments: OPERA and MINOS. This main chain maybe summarized in the simplest way as follows (more details are hidden within table caption). In first Table 1, we estimated: \\
    \begin{enumerate}
	\item   the chord distance among the sources (Cern or FNAL) and the most distant neutrino detectors (SK or ICECUBE);\\
    \item  point 1) leading to the corresponding tuned energy for each chord distance to make neutrino oscillation (first in vacuum and later on in Earth matter) detectable in DeepCore, as well as vanishing in CPT conserved model (but not in the violated one). These tuned energies are indeed found to be (for matter realistic case), around twenty GeVs;\\
   \item   the ratio among these longest base distances and OPERA one. OPERA has been used to calibrate the experiment once we correctly were able to evaluate their performance. From these ratio we estimated  the neutrino flux dilution factor $(\frac{d'}{d})^{2}$;\\
  \item    for each energy and source-detector we remind each  mass detector;\\
   \item   for each tuned energy we estimated the corresponding mass detector and considered the $\nu_{\mu}$ event rate both for charged  and neutral current inside the volume;\\
  \item    we re-scale the event rate for each year and each kiloton of any detector, assuming no oscillations.\
\end{enumerate}

In the following Table 2 we kept care of the beam bending at each tuned energy considered in previous table and steps, (see Fig.\ref{Tunnel-CERN}, Fig.\ref{Tunnel-Fermi}). Namely we estimated:\\
       \begin{enumerate}
     \item each beam bending radius, the Larmor radius $R_{L}$, assuming a nominal one Tesla corn as magnetic bending lens.\\
     \item From point 1) we evaluated each corresponding arc length toward each (SK or IceCube) target;\\
     \item a complementary $\pi$ decay in flight in a shorter tunnel able to allow only $20\%$ of the pion decay. These length  avoid a too deep and costly tunnel. We also assume a quasi monochromatic beam  that reduce at least by a factor $50\%$  the neutrino flux. Therefore (with a $20$\% tunnel size) these suppressions are leading to a $10\%$ efficiency with respect to OPERA experiment. Moreover, because of  redundancy, we also considered the minimal configuration with a tunnel as small as $5\%$  of OPERA one and a quasi-monochromatic beam  as small as factor $20\%$ of OPERA, is anyway offering a $1\%$ whose ability to reveal CPT violation and tau and anti-tau appearance is remarkable;\\
     \item the vertical depth corresponding   to such a limited flight tunnel.
\end{enumerate}

\begin{flushleft}
\begin{table*}[htbp]
\begin{center}
\begin{small}
\begin{tabular}{lccccccc}
\hline
\hline
&&&&&&\\
\multicolumn{8}{c}{Muon Neutrino beam events by}\\
\multicolumn{8}{c}{3.5 $\cdot 10^{19}$ proton on target(p.o.t) a year from CERN or FNAL }\\
&&&&&&\\
\hline
&&&&&&\\
Baseline & $distance$ &  $E_{\nu}$ & $\left(\frac{L'}{L}\right)^2$ & Mass detector  &  $N_{ev_{CC+NC}} $ & $N_{ev \mu_{CC}} $ no osc.  & $N_{ev \mu_{CC}} $ no osc. \\
&(km) & (GeV)  & & $kton$ & $M_{in}^{-1}$$ year^{-1}$ & $kton^{-1}$$ year^{-1}$ &  $ year^{-1}$\\
\hline
&&&&&&\\
CERN--OPERA&$        L =732$ &    $17$  &   1 &      $ 1.2 $  & 3500 &    $ 2370$ & 2847     \\
CERN--SK&$            L' =8737$& $15.8$  & $142.5$ & $ 22.5$ & $ 398.5 $ & $14.41$&  324   \\
Fermilab--SK&$       L' =9140$& $16.5$  &    $155.9$ & $ 22.5$ & $ 420$ & $ 15.18$ &  341   \\
CERN-IceCube&$       L' =11812$& $21.8$  & $260.4$ & $ 4800$ & $114750 $ & $19.45$ &  93343  \\
Fermilab--IceCube&  $L' =11623$& $21.4$ &  $252.1$ & $ 4800$ & $115500$ & $19.57$   &  93951   \\
&&&&&&\\
\hline
\end{tabular}
\end{small}
\end{center}
\caption{ Source detector distances, tuned energies, flux dilution, event rate }\label{f}

%


\begin{center}
\begin{small}
\begin{tabular}{lccccccccc}
\hline
\hline
Baseline & $ E_{\nu}$ & $E_{\pi}$  &    Angle     &   $R $          &$L_{arc}$  & Arc Depth \\
         &(GeV)       &  (GeV)     &    degrees   &       (m)       &    (m)    &  (m)     \\
\hline
&&&&&&&&\\
CERN--SK&            $15.8$& $37.9$  &   43.19°     &$126.5$ & $95.3$  & $ 34.2$      \\
Fermilab--SK&        $16.5$  & $39.8$  &    45.77°    &$132.7$ & $106$   & $ 40  $      \\
CERN-IceCube&        $21.8$& $51.3$  &    67.82°    &$171$   & $202.3$ & $ 106.3$     \\
Fermilab--IceCube&   $21.4$& $50.8$  &    65.67°    &$169.4$ & $194.2$ & $ 99.5$      \\
\hline
\end{tabular}\\
\begin{tabular}{lccccccccc}
\hline
\hline
Baseline &  Tunnel                   & Tunnel                   & Tunnel           &   Tunnel        \\
  & length $L_{20\%}$ &  depth $H_{20\%}$       & length $L_{5\%}$ &   depth $H_{5\%}$        \\
& (m) & (m) &  (m)&  (m) \\
\hline
&&&&&&&&\\
CERN--SK&              202  & $138$ & $50$ &34   \\
Fermilab--SK&         222  & $159$ & $55$ &40   \\
CERN-IceCube&          369  & $342$ & $92$ &85    \\
Fermilab--IceCube&    362  & $330$ & $90$ &82    \\
\hline
\end{tabular}\\
\caption{Beaming, bending and tunnel parameters. Above: final neutrino energy and parent pion energy, bending angle, radius of curvature for 1 T magnetic field, beam bending arc length and depth. Below: decay tunnel length and depth in two economic scenario, that are considering two shorter  pion decay tunnel length (respect OPERA one -1 km-):
the first one is $20\%$ length reduction and the second one is $5\%$ length reduction.}\label{tunnell}
\end{small}
\end{center}
\end{table*}
\end{flushleft}

\begin{figure}
\begin{center}
\includegraphics[angle=0,scale=.33]{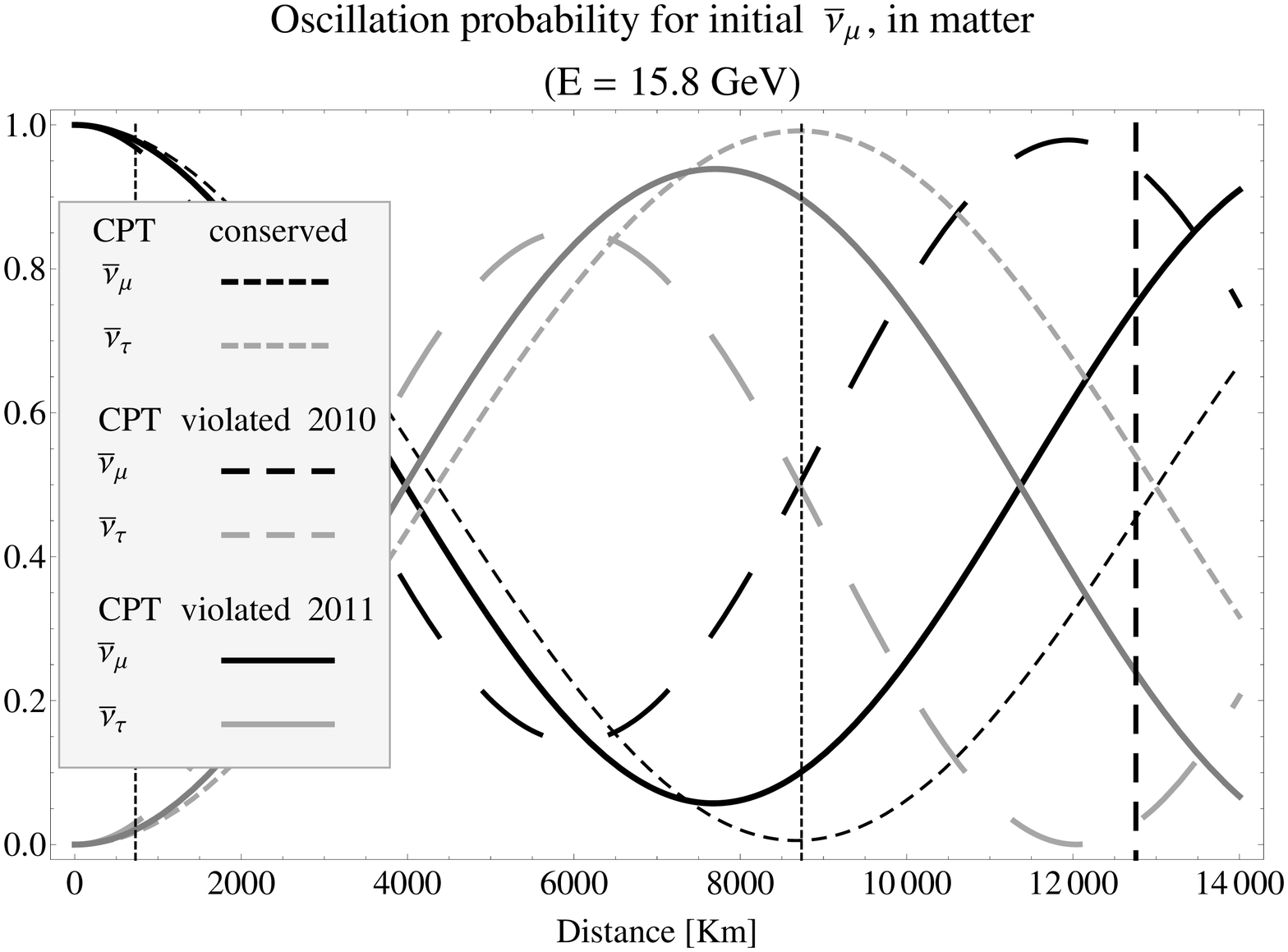}
\includegraphics[angle=0,scale=.33]{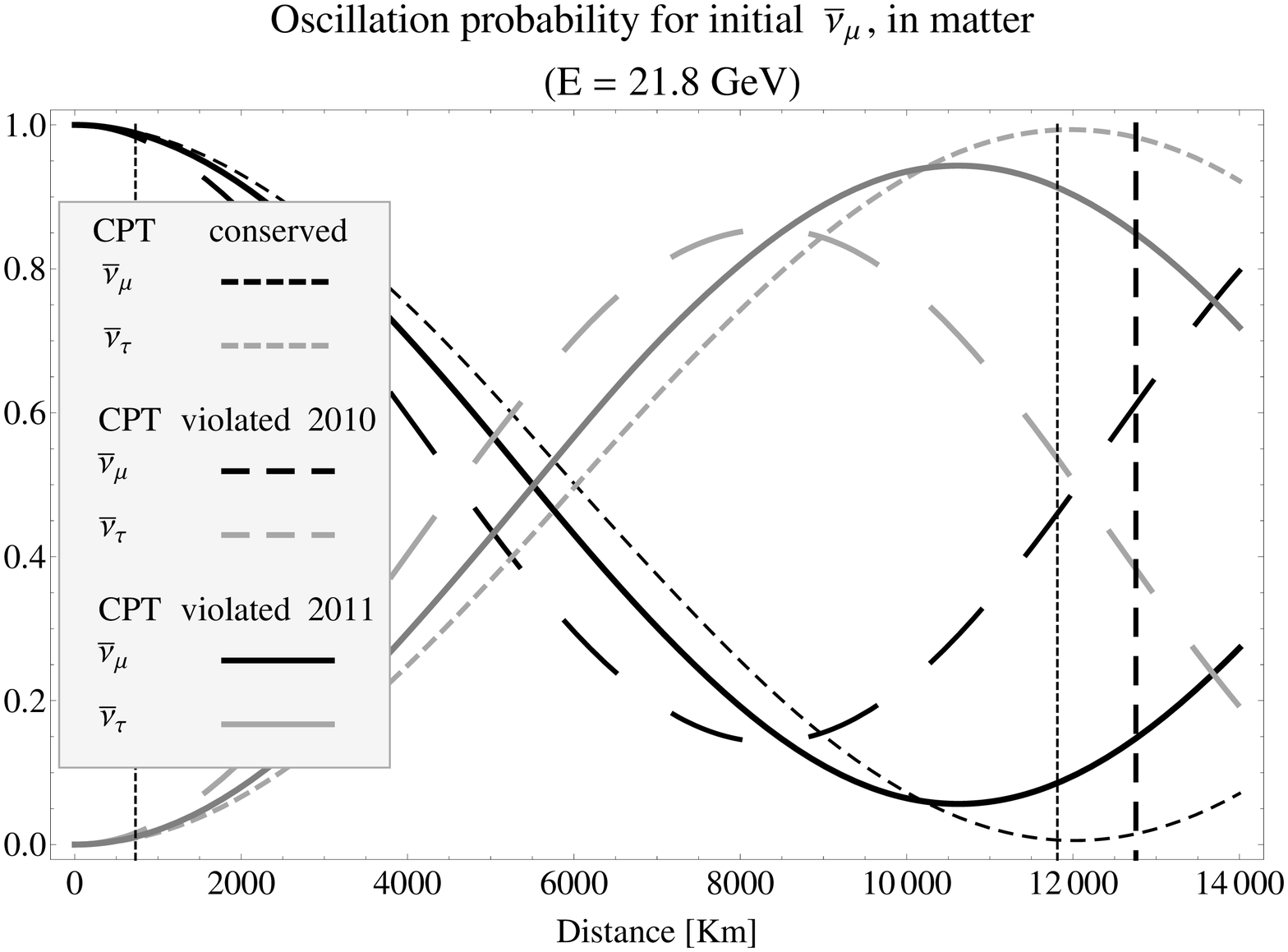}
\caption{The $\bar{\nu}_{\mu}$ survival and oscillation probability (in three flavor case)  for various CPT cases. Neutrino beam sent from CERN (to IceCube or SuperK) at tuned energy for  muon neutrino total disappearance, in CPT conserved are considered. As in fig. \ref{02} three cases are considered.  The light vertical dotted lines stand for the CERN-SuperK ($E = 15.8$ GeV) or CERN-IceCube ($E = 21.8$ GeV) distances, (also shown the CERN-Opera distance on the extreme left), while heavy dashed line show the Earth diameter. }\label{05}
\end{center}
\end{figure}

\begin{figure}
\begin{center}
\includegraphics[angle=0,scale=.35]{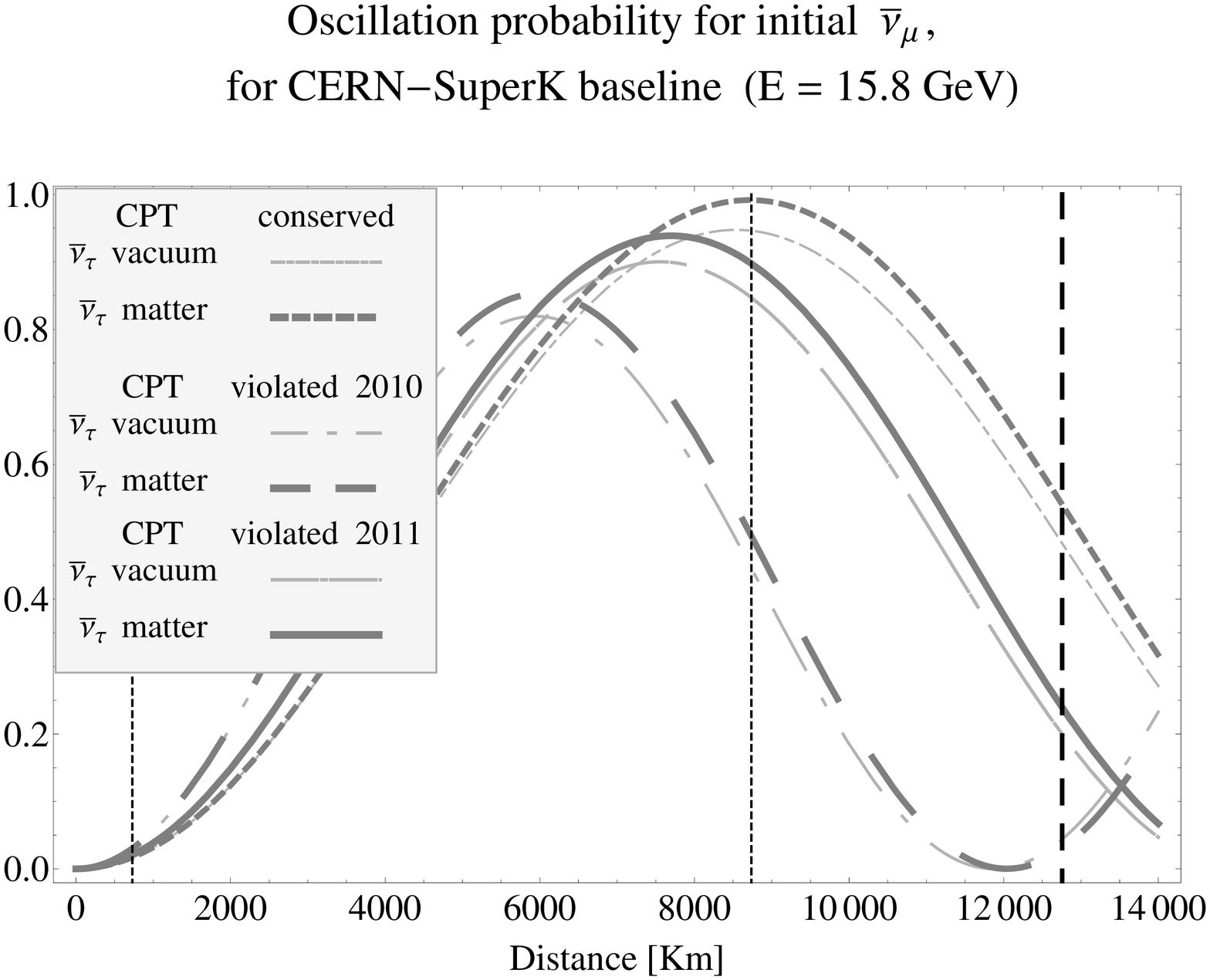}
\includegraphics[angle=0,scale=.35]{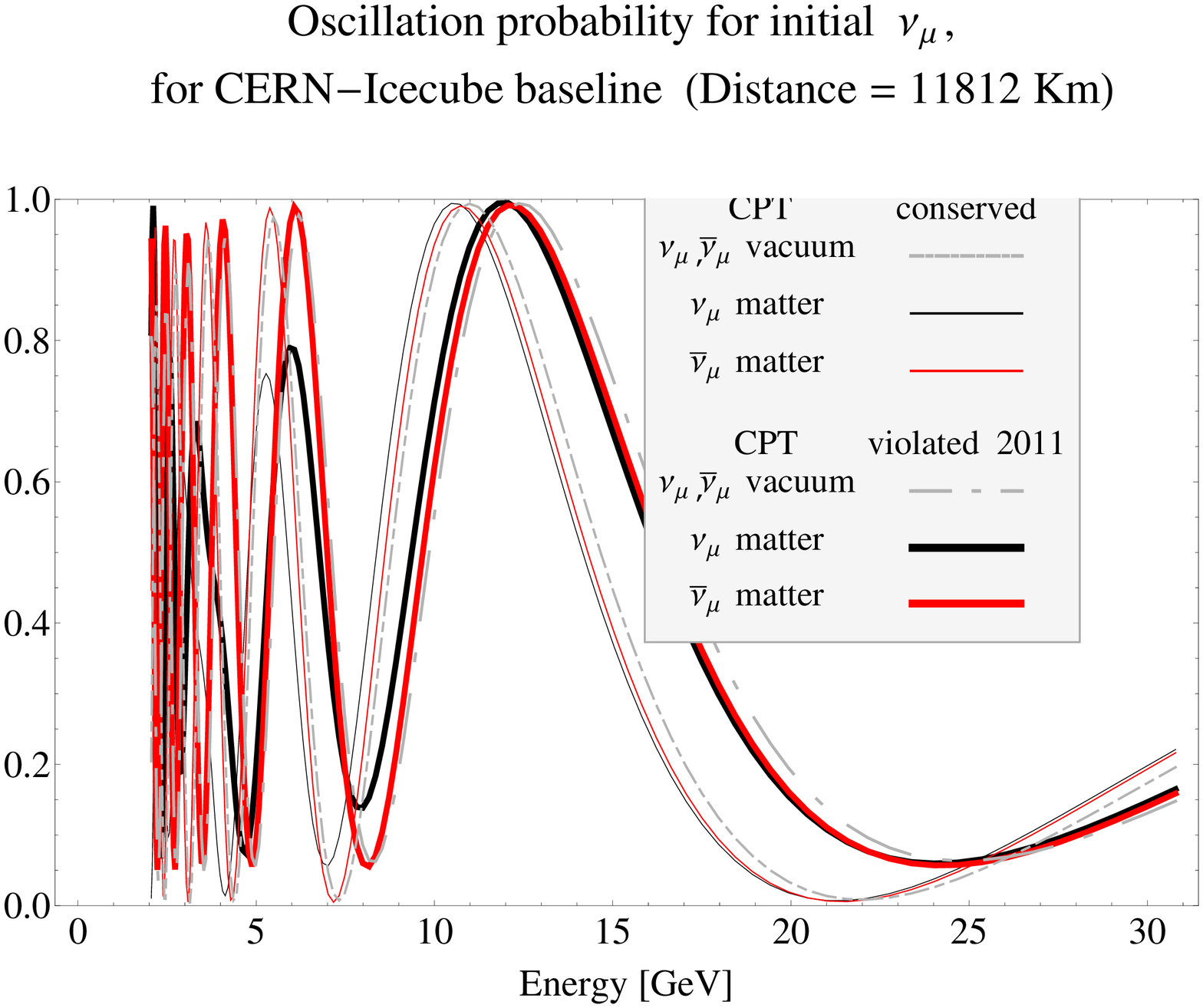}
\caption{Above: anti-neutrino tau appearance by oscillation probability ($\overline{\nu}_{\mu}\rightarrow\overline{\nu}_{\tau} $) for CERN-SK baseline tuned energy. The three cases are considered, both in vacuum and in earth matter. The vertical light dotted line stand for the CERN-SK distance.
Below: muon neutrino and anti-neutrino survival probability ($\nu_{\mu}\rightarrow\nu_{\mu} $). It is shown the difference between neutrino and anti-neutrino across matter (because a tiny asymmetry in MSW term, see Appendix), while such asymmetry in CPT conserved case is absent  in vacuum; moreover such discrepancy between matter and anti-matter are negligible in our energy window of interest.
The FNAL-SK and FNAL-ICECUBE oscillation are very comparable to the Cern ones because very similar distances. This twin coincidences are very remarkable  indeed. Note that while crossing the Earth here and later we did not consider just (as most other authors) the average Earth density  ( respectively $\rho= 4.5$ $g/cm^{3}$ for  "near" SK; $\rho= 7.2$ $g/cm^{3}$  for  "far" IceCube DeepCore), but the exact variable matter profile and we did estimate the mixing  step by step considering the influence MSW matter along inside the Earth. Nevertheless these vacuum and matter cases do not differ much among themselves for energies higher than ten GeVs. }\label{06}
\end{center}
\end{figure}

Note that at few GeV energy there is a remarkable deviation between vacuum and our exact matter neutrino mixing.
The third table shows the event rate a year for $\nu_{\mu}$, $\bar{\nu}_{\mu}$, as in case of no oscillation. The first column keep the estimate of last column of Table 1, then we:\\
     \begin{itemize}
     \item Re-scale the event result for each internal mass with no oscillation.\\
     \item Evaluate the external muon rate by  corresponding detector mass environmental and density around, with no oscillation considered.\\
     \item Combine the whole rate taking into of internal and external mass.\\

       From here in next fourth table,  keeping care of the matter influence:

     \item we did considered the oscillating probability (theoretically vanishing in CPT conserved model) also taking care of the neutrino energy spread (by $\frac{ \Delta E}{E} = 10\%$) assumed from the source.\\
     \item We estimated the same neutrino event rate to appear after the crossing and mixing along the Earth chord.\\
     \item We conclude these estimates for anti-neutrinos both in CPT conserved and violated case.\\
     \end{itemize}

 In the fifth table as above we assumed a smaller ($10\%$), ($1\%$) configuration regarding the muon neutrino survival and rate in a recording year.
 As above in the sixth table we did estimate the same event rate in DeepCore for a $100\%$ OPERA experiment, assuming the pessimistic energy spread as large as $\frac{ \Delta E}{E} = 20\%$. Finally on the Table 7 we did estimate the same event rate in DeepCore for a minimal  ($10\%$), ($1\%$) OPERA-like experiment, assuming as before a pessimistic energy spread as large as $\frac{ \Delta E}{E} = 20\%$. In this table 7, even within the worst minimal scenario we foresee hundreds of muon events able to clearly disentangle CPT conserved from CPT violated case at $6.1-6.3\  \sigma$ level .
Next tables $8-9$ reproduce in similar way the prediction for  $\tau$ and ${\bar{\tau}}$ appearance in matter, keeping care of the Earth profile. Once again  because of the extreme statistical power we considered even the most economic scenario at $1\%$ of a beaming (respect a $100\%$ tunnel and flux as OPERA experiment). We may reveal  the tau appearance by their showering rate over a comparable Neutral Current showering noise. The effect is not as in OPERA event diagnostic, but the statistical signature of tau (and anti tau) showering is solid. In the last row of Table $7$ we just  mention also a related (to Table $9$, tau appearance) muon birth as a secondary in tau decay at $17.4\%$ channel level and at low energy ($\sim 7$) GeV. Conclusion and Discussions summarize these Table results.
\begin{figure}
\begin{center}
\includegraphics[angle=0,scale=0.37]{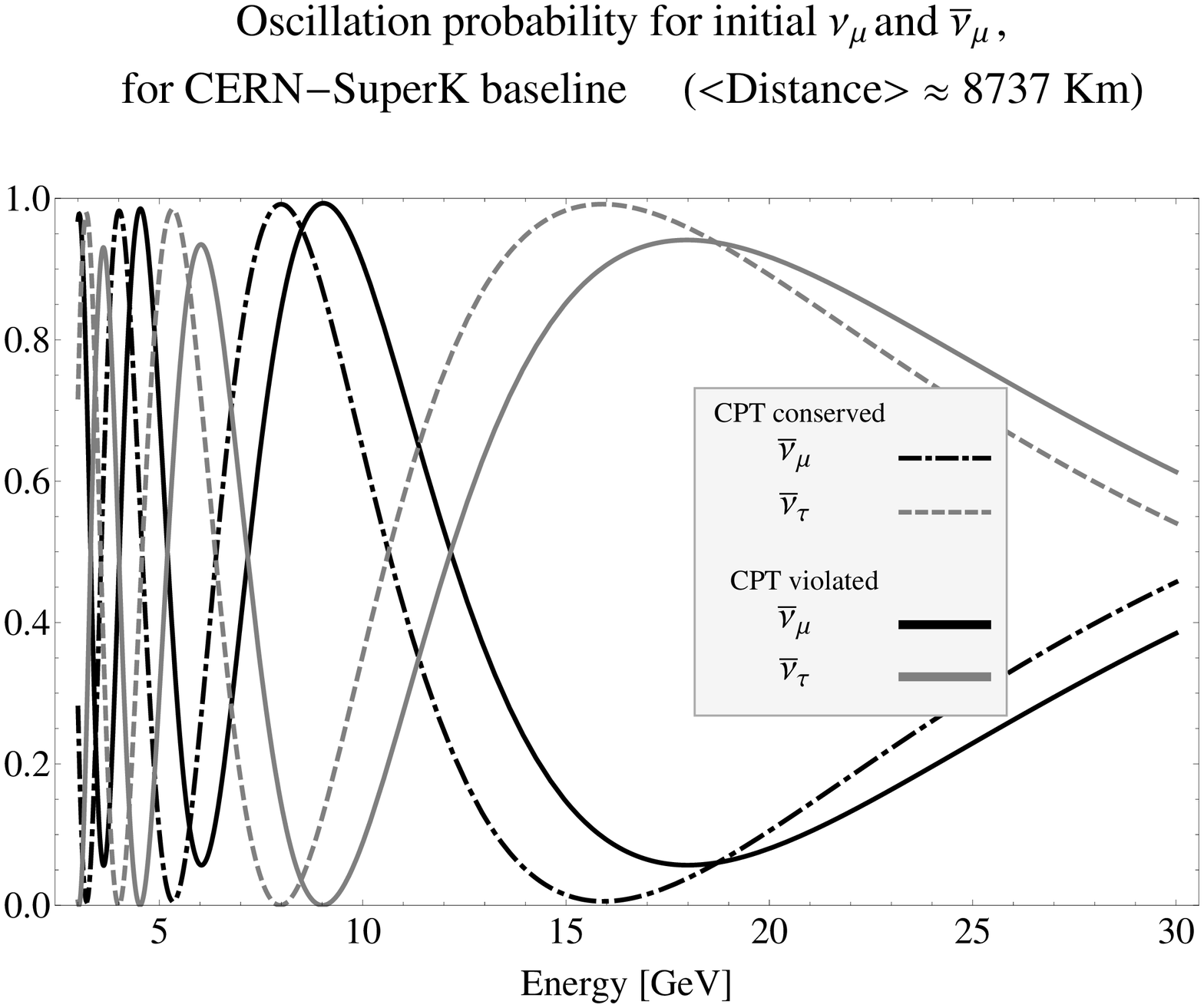}
\caption{Muon anti-neutrino oscillation probability  ($\overline{\nu}_{\mu}\rightarrow\overline {\nu}_{\tau}$) and survival probability ($\overline{\nu}_{\mu}\rightarrow\overline {\nu}_{\mu}$) in last CPT violated case \cite{111}, keeping care of the matter influence, for Cern-SK baseline.  Here we did consider an ideal  monochromatic energy neutrino beam, while in next figures we will address to a realistic smeared energetic beam. Already here we note the mild discrepancy of the neutrino muon suppression at its minimum (15.8 GeV)  between CPT conserved and violated case. However this is not large enough to disentangle the eventual broken symmetry in this baseline, because of the small SuperK detector, respect to IceCube, in the following figure. }\label{081}
\includegraphics[angle=0,scale=0.37]{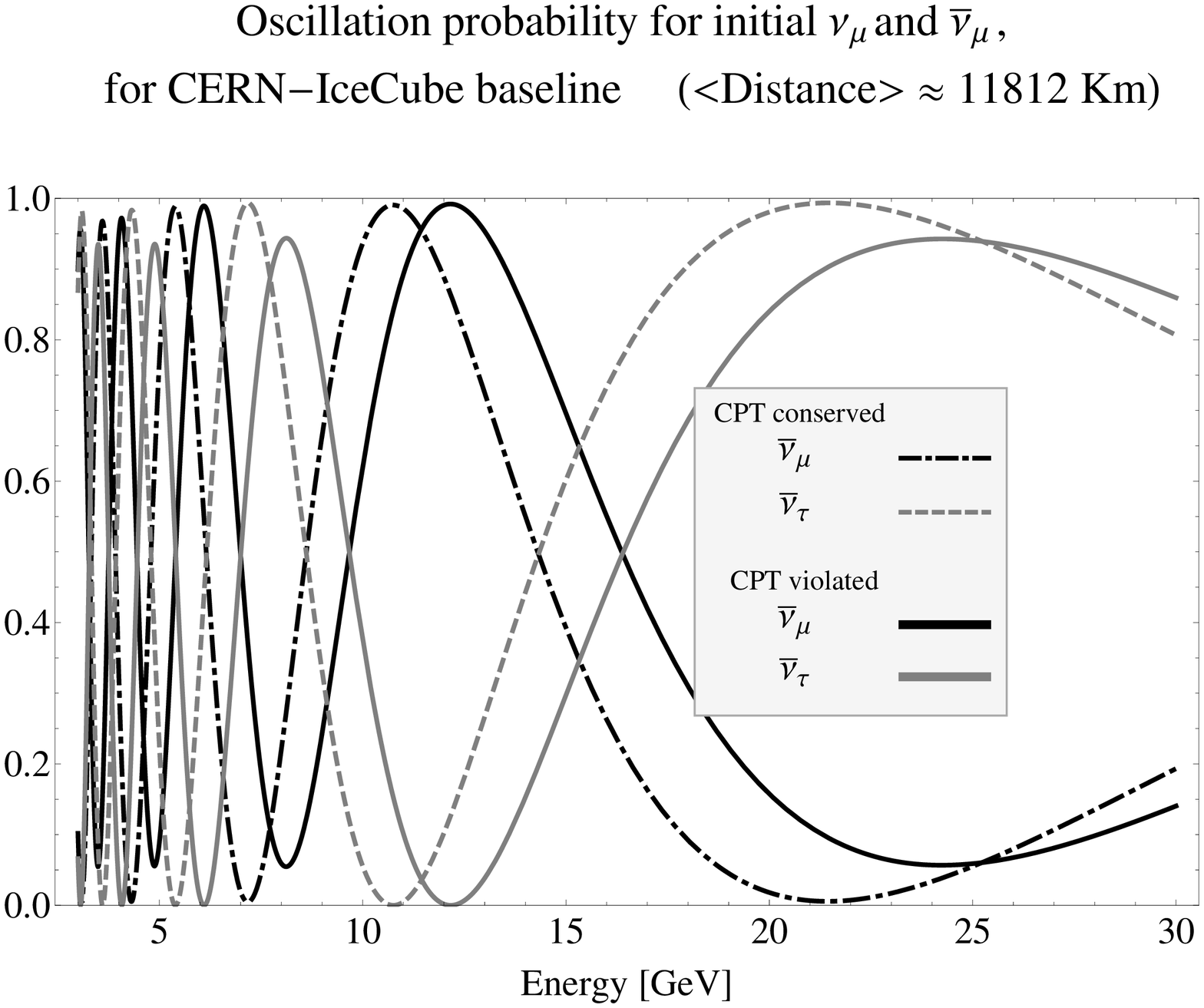}
\caption{Muon anti-neutrino oscillation probability  ($\overline{\nu}_{\mu}\rightarrow\overline {\nu}_{\tau}$) and survival probability ($\overline{\nu}_{\mu}\rightarrow\overline {\nu}_{\mu}$) in last CPT violated case \cite{111}, keeping care of the matter influence, for Cern-ICECUBE baseline. Here we did consider an ideal  monochromatic energy neutrino beam, while in next figures we will address to a realistic smeared energetic beam. Here we note that the mild discrepancy of the neutrino muon suppression at its minimum (21.8 GeV)  between CPT conserved and violated case, is large enough to disentangle the eventual broken symmetry, even at $1\%$ OPERA like experiment (by flux and tunnel length reduction). }\label{082}
\end{center}
\end{figure}


\subsection{Notes on beaming Neutrino across the Earth }
In order to obtain an approximate neutrino energy near 20 GeV from pion decay some issues arise.
The parent pion energy, itself originated from a 400 GeV proton beam, has to be 50 GeV at least, which is a factor $\sim2.3$ greater than the required neutrino energy. Indeed accordingly to the relativistic relation, we find:
$
E_{\nu }^{\max }=\frac{m_{\pi }^2-m_{\mu }^2}{2 m_{\pi }^2}\left(E_{\pi }+p_{\pi }\right)\approx \left(1-\frac{m_{\mu }^2}{m_{\pi}^2}\right)E_{\pi }= 0.427 E_{\pi }$

This maximum neutrino energy occur at forward decay angle in pion rest frame and, after boosting, the highest part of energy distribution is selected toward detector direction. The main relativistic beaming angle is quite small: $\Delta \theta \leq \frac{m_{\pi}}{E_{\pi}}\sim \frac{1}{350} \left( \frac{E_{\pi}}{53 GeV}\right)^{-1} $. Nevertheless within such a small angle a large $\phi_{\nu}$ fraction is contained. The spread of such a beam from Cern to IceCube produce a flux whose diameter $d\leq 2 R_{\oplus}\sin{68°} \cdot \frac{m_{\pi}}{E_{\pi}} \simeq 34 Km$.
 To increase the beam monochromaticity, the 50 GeV pion beam is bent with 1 Tesla magnetic field, as shown in table \ref{tunnell}.
 For obvious reasons we didn't take in account the additional $K^+ K^-$ decay, mostly because larger masses, lower lorentz factor and beaming, three body decay.

\begin{figure}
\begin{center}
\includegraphics[angle=0,scale=0.4]{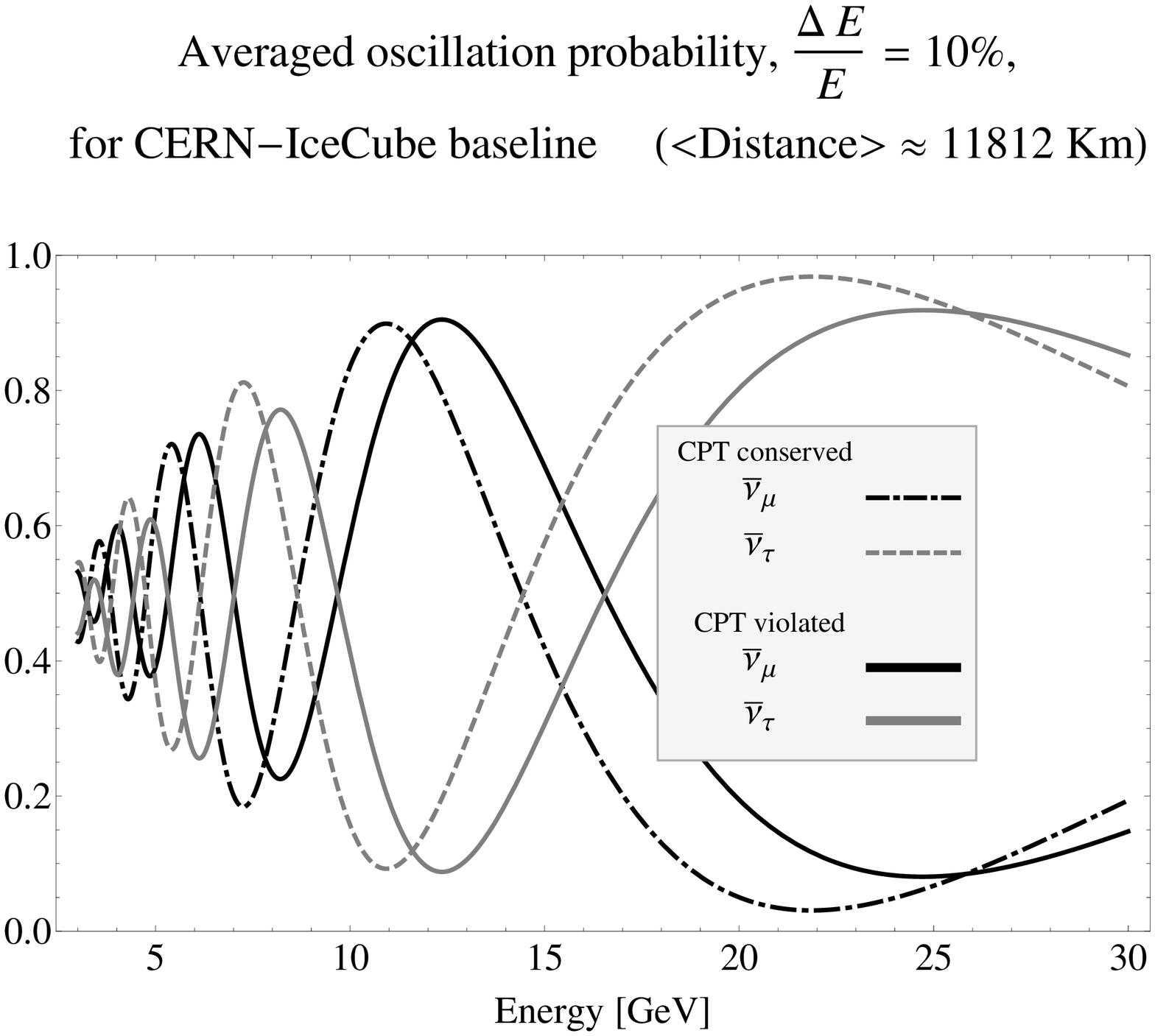}
\caption{Averaged oscillation and survival probability, showing the average conversion smeared by neutrino energy spectra, at $ \frac{\Delta E}{E}$$= \pm 10$\%. The survival probability for $\bar{\nu}_{\mu}$
in CPT conserved and violated cases is less sharp than in the monochromatic scenario ($P_{\mu \mu} = 0.03$ and $0.13$, respectively). However as shown in table n.5, the event rate even in the framework of a minimal $1\%$ Opera-Like experiment allows in one year to reach $6.6 \sigma$ signature for an hypothetical CPT violation  within earliest (2010) CPT deviations and present (2012) MINOS bounds. } \label{deltaE10}
\includegraphics[angle=0,scale=0.4]{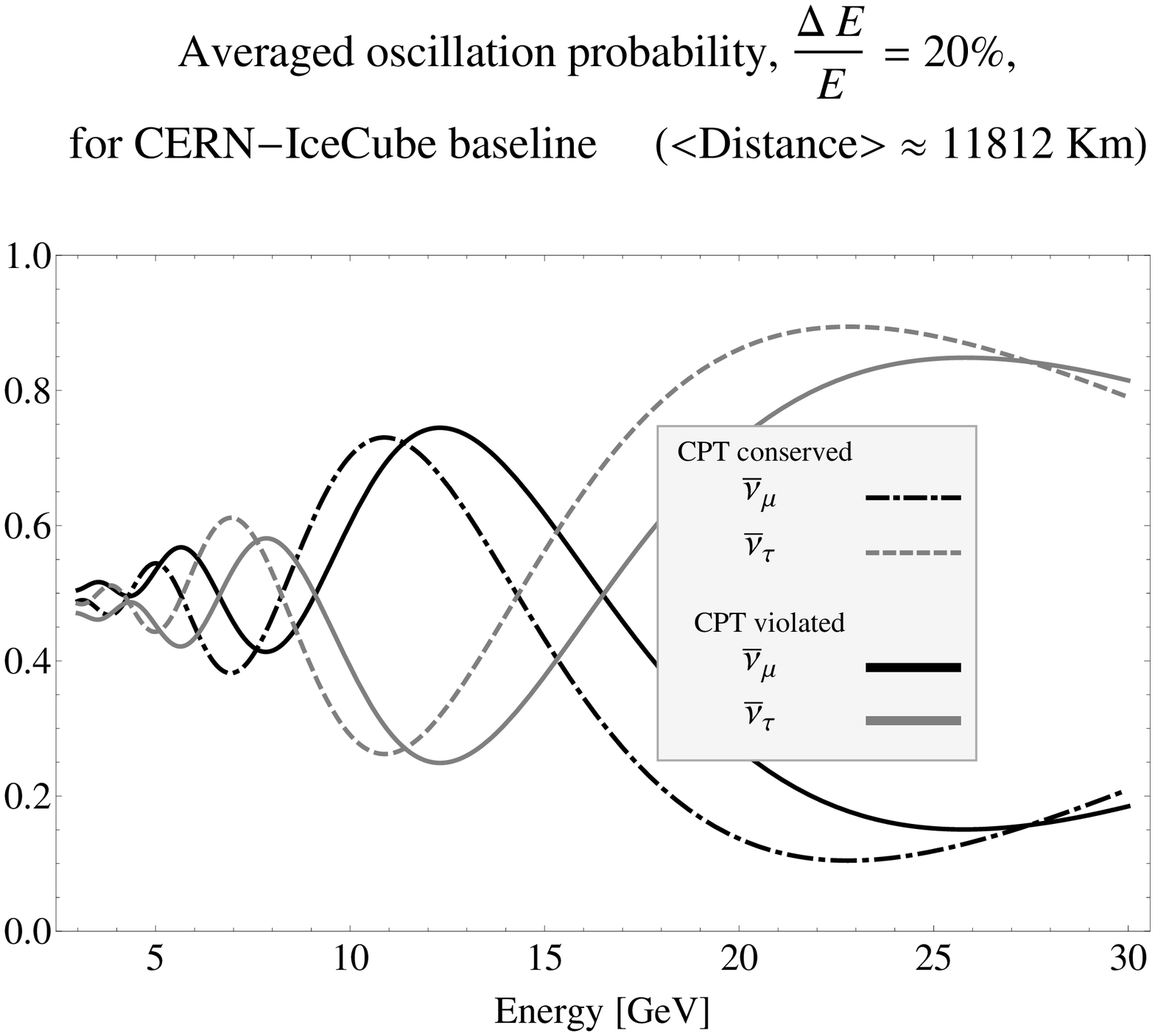}
\caption{Averaged oscillation and survival probability, showing the average conversion smeared by neutrino energy spectra, at $ \frac{\Delta E}{E}$$= \pm 20$\%. The survival probability for $\bar{\nu}_{\mu}$ in CPT conserved and violated cases is less sharp than in the monochromatic scenario ($P_{\mu \mu} = 0.13$ and $0.26$, respectively). However as shown in table n.7, the event rate even in the framework of a minimal $1\%$ Opera-Like experiment allows in one year to reach $6.1 - 6.3\ \sigma$ signature for an hypothetical CPT violation  within earliest (2010) CPT deviations and present (2012) MINOS bounds.  }\label{deltaE20}
\end{center}
\end{figure}

\begin{table*}[htbp]
\begin{center}
\caption{$\nu_{\mu}$ and $\bar{\nu}_{\mu}$ appearance: event rate in conserved and in violated CPT
parameters (100\% tunnel length comparable with
OPERA).
The events are relative to one year data taking, as they would be without neutrino oscillation.
From the first column: $\nu_{\mu}$ energy at maximum conversion into $\nu_{\tau}$, under CPT conserved parameters; the number of $\nu_{\mu}$
events inside detector as already shown in table 1; number of events whose track starts outside detector; sum of previous $\nu_{\mu}$ events;
total events for $\bar{\nu}_{\mu}$, taking into account half cross section. }\label{f3}
\begin{tabular}{lccccc}
\hline
\hline
Baseline & $ E_{\nu}$ & $N_{\nu_{\mu}}$ no osc. & $N_{\nu_{\mu}}$ no osc. & $N_{\nu_{\mu}}$  no osc. & $N_{\bar{\nu}_{\mu}}$ no osc.\\
    &    &born in detector  &   born outside  &  total    &  total \\
    &  GeV  &year$^{-1}$       & year$^{-1}$           &year$^{-1}$        &   year$^{-1}$    \\
\hline
CERN--OPERA   & 17  & $2847$  & $18278$     & $21125$      & $10562$  \\
&&&&&\\
CERN--SK      & 15.8    & $324$   & $1621$  & $1945$   & $972$     \\
Fermilab--SK  & 16.5    & $342$   & $1708$  & $2050$   & $1025$    \\
CERN--IceCube &21.8     & $93343$ & $23336$ & $116679$ & $58340$    \\
Fermilab--IceCube &21.4 & $93951$ & $23488$ & $117439$ & $58720$    \\
\hline
\end{tabular}\\

\caption{$\nu_{\mu}$ and $\bar{\nu}_{\mu}$ appearance: event rate in conserved and in violated CPT
parameters (100\% tunnel length comparable with OPERA). The events are relative to one year data taking, taking into account neutrino oscillation. The survival probability for $\nu_{\mu}$, considering $\frac{\Delta E}{E} = 10 \%$ energy spread, is shown, under CPT conserved parameters, and the relative events for $\nu_{\mu}$ and $\bar{\nu}_{\mu}$; then the survival probability for  CPT violated parameters (2012) is shown, with relative $\bar{\nu}_{\mu}$ events only.}\label{lf}
\begin{tabular}{lccccc}
\hline
\hline
Baseline  & $\langle P_{\mu \mu} \rangle$  & $N_{\nu_{\mu}}$ after osc.  & $N_{\bar{\nu}_{\mu}}$ after osc. & $\langle P_{\bar{\mu} \bar{\mu}}\rangle$  & $N_{\bar{\nu}_{\mu}}$ after osc. \\
          & $\mathbf{\frac{\Delta E}{E} = 10 \%}$   & year$^{-1}$ &  year$^{-1}$ & $\mathbf{\frac{\Delta E}{E} = 10 \%}$ &  year$^{-1}$ \\
        &  $\exists$ CPT    &   $\exists$ CPT     &   $\exists$ CPT      &  $\nexists$ CPT   &     $\nexists$ CPT      \\
\hline
CERN--OPERA       & 0.985 &  20494  & 10247 &  0.972 &   10125 \\
&&&&&\\
CERN--SK          & 0.029 &  56    &  28   &    0.115     &  112  \\
Fermilab--SK      & 0.029 &  59    &  30   &    0.114     &  117 \\
CERN--IceCube     & 0.031 &  3660  &  1830 &    0.13     &  7584 \\
Fermilab--IceCube & 0.031 &  3641  &  1820 &    0.129    &  7575\\
\hline

\end{tabular}
\end{center}
\end{table*}

\begin{table*}[htbp]
\begin{center}
\caption{$\nu_{\mu}$ and $\bar{\nu}_{\mu}$ event rates considering reduced size experiment, and $\frac{\Delta E}{E} = 10 \%$ energy spread.
The first row shows reduced event numbers to $10\%$, because of shorter decay tunnel ($50\%$ respect OPERA one) and reduced $\pi$ flux intensity for narrow neutrino spectra ($20\%$ respect OPERA flux) due to suggested setup for pion bending magnetic field.
The second row shows a very economic scenario, where event rate is reduced to $1\%$, having considered $20\%$ decay tunnel, and $5\%$  $\pi$ flux intensity. In the last row, $N_{\mu_{\tau}}$ and $N_{\bar{\mu}_{\tau}}$ are the $\mu$, $\bar{\mu}$ by $\tau$ and $\bar{\tau}$ decay.} \label{f5}
\begin{small}
\begin{tabular}{lcccc}
\hline
\hline
\textbf{10\% } & $ N_{\nu_{\mu}}$ after osc.  & $N_{\bar{\nu}_{\mu}}$ after osc. & $N_{\bar{\nu}_{\mu}}$ after osc. & Statistical \\
        & year$^{-1}$  &   year$^{-1}$     &    year$^{-1}$  & Significance \\
Baseline&  $\exists$ CPT &   $\exists$ CPT   &  $\nexists$ CPT  & $\sigma$ \\
\hline
CERN--SK        &    $6$       & $\mathbf{3} $    & $\mathbf{11 \pm3}$      & 2.5  \\
Fermilab--SK    &    $6$       & $\mathbf{3}$     & $\mathbf{12 \pm3}$      & 2.5   \\
CERN--IceCube   &    $366$     & $\mathbf{183} $   & $\mathbf{758 \pm27}$  & 21   \\
Fermilab--IceCube&   $364$     & $\mathbf{182} $   & $\mathbf{757 \pm27}$  & 21   \\
\hline

 \textbf{1\% }&  $ N_{\nu_{\mu}}$ after osc. & $N_{\bar{\nu}_{\mu}}$ after osc. & $N_{\bar{\nu}_{\mu}}$ after osc. & Statistical \\
        & year$^{-1}$  &   year$^{-1}$     &    year$^{-1}$  & Significance \\
Baseline&  $\exists$ CPT &   $\exists$ CPT   &  $\nexists$ CPT  & $\sigma$ \\
\hline
CERN--SK       &     $0.6$     & $\mathbf{0.3}$     & $\mathbf{1.1\pm1}$  & 0.8  \\
Fermilab--SK    &   $0.6$      & $\mathbf{0.3}$       & $\mathbf{1.2\pm1}$  & 0.8  \\
CERN--IceCube  &     $37$     & $\mathbf{18}$    & $\mathbf{76\pm9}$  & 6.6   \\
Fermilab--IceCube&   $36$     & $\mathbf{18}$    & $\mathbf{76\pm9}$  & 6.6   \\
\hline
 \textbf{1\% }&   $\langle E_{\mu_{\tau}} \rangle $(Gev) &      $N_{\mu_{\tau}}^{*} $ & $ N_{\bar{\mu}_{\tau}}^{*}$   &  $N_{\bar{\mu}_{\tau}}^{*}$   \\
Baseline&  &   CPT conserved & CPT conserved &  CPT violated\\
\hline
CERN--SK       &     $5.4$     &  0.2 &   0.1     & 0.5 \\
Fermilab--SK    &    $5.6 $      &    0.2 & 0.1     & 0.5 \\
CERN--IceCube  &     $7.3$     &$\mathbf{ 62} $ &  $ \mathbf{31}$    &  $\mathbf{ 13.5 } $   \\
Fermilab--IceCube&   $7.2$      &   $\mathbf{60}$ & $\mathbf{30} $     &   $\mathbf{ 12.5 } $  \\
\hline
\multicolumn{5}{l}{* Considering a $17.4 \% $ branching ratio.}\\
\multicolumn{5}{l}{We remind that the ${\mu_{\tau}}$ energy is nearly $\simeq \frac{1}{3}$ of primary tau.}\\
\multicolumn{5}{l}{ The name ${\mu_{\tau}}$ (in \cite{-7}), has been much later renamed Tautsie-pop in \cite{-8}. }\\
\end{tabular}
\end{small}
\end{center}
\end{table*}

\begin{table*}[htbp]
\begin{center}
\caption{$\nu_{\mu}$ and $\bar{\nu}_{\mu}$ appearance: event rate in conserved and in violated CPT
parameters (100\% tunnel length comparable with OPERA). The events are relative to one year data taking, taking into account neutrino oscillation. The survival probability for $\nu_{\mu}$, considering $\frac{\Delta E}{E} = 20 \%$  energy spread is shown, under CPT conserved parameters, and the relative events for $\nu_{\mu}$ and $\bar{\nu}_{\mu}$;
then the survival probability for  CPT violated parameters (2012) is shown, with relative $\bar{\nu}_{\mu}$ events only.}\label{f6}
\begin{tabular}{lccccc}
\hline
\hline
Baseline  & $\langle P_{\mu \mu} \rangle$  & $N_{\nu_{\mu}}$ after osc.  & $N_{\bar{\nu}_{\mu}}$ after osc. & $\langle P_{\bar{\mu} \bar{\mu}}\rangle$  & $N_{\bar{\nu}_{\mu}}$ after osc. \\
          & $\mathbf{\frac{\Delta E}{E} = 20 \%}$   & year$^{-1}$ &  year$^{-1}$ & $\mathbf{\frac{\Delta E}{E} = 20 \%}$ &  year$^{-1}$ \\
        &  $\exists$ CPT    &   $\exists$ CPT     &   $\exists$ CPT      &  $\nexists$ CPT   &     $\nexists$ CPT      \\
\hline
CERN--OPERA       & 0.985 &  20494  & 10247 &  0.972 &   10125 \\
&&&&&\\
CERN--SK          & 0.096 &  187    &  93   &    0.18    &  175   \\
Fermilab--SK      & 0.096 &  197    &  98   &    0.179   &  183   \\
CERN--IceCube     & 0.13 &  12415  &  6207 &    0.26     &  15168 \\
Fermilab--IceCube & 0.129 &  12496  &  6248 &    0.263   &  15443 \\
\hline
\end{tabular}
\end{center}
\end{table*}

\begin{table*}[htbp]
\begin{center}
\caption{$\nu_{\mu}$ and $\bar{\nu}_{\mu}$ event rates considering reduced size experiment, and $\frac{\Delta E}{E} = 20 \%$ energy spread.
The first row shows reduced event numbers to $10\%$, because of shorter decay tunnel ($50\%$ respect OPERA one) and reduced $\pi$ flux intensity for narrow neutrino spectra ($20\%$ respect OPERA flux) due to suggested setup for pion bending magnetic field.
The second row shows a very economic scenario, where event rate is reduced to $1\%$, having considered $20\%$ decay tunnel, and $5\%$  $\pi$ flux intensity.
In the last row, $N_{\mu_{\tau}}$ and $N_{\bar{\mu}_{\tau}}$ are the $\mu$, $\bar{\mu}$ by $\tau$ and $\bar{\tau}$ decay.} \label{f7}
\begin{small}
\begin{tabular}{lcccc}
\hline
\hline
 & $ N_{\nu_{\mu}}$ after osc.  & $N_{\bar{\nu}_{\mu}}$ after osc. & $N_{\bar{\nu}_{\mu}}$ after osc. & Statistical \\
        & year$^{-1}$  &   year$^{-1}$     &    year$^{-1}$  & Significance \\
Baseline&  $\exists$ CPT &   $\exists$ CPT   &  $\nexists$ CPT  & $\sigma$ \\
\hline
CERN--SK        &    $19$       & $\mathbf{9} $    & $\mathbf{18 \pm4}$      & 2    \\
Fermilab--SK    &    $20$       & $\mathbf{10}$     & $\mathbf{18 \pm4}$      & 2   \\
CERN--IceCube   &    $1516$     & $\mathbf{758} $   & $\mathbf{1517 \pm39}$  & 19   \\
Fermilab--IceCube&   $1515$     & $\mathbf{757} $   & $\mathbf{1544 \pm39}$  & 20   \\
\hline

 \textbf{1\% }&  $ N_{\nu_{\mu}}$ after osc. & $N_{\bar{\nu}_{\mu}}$ after osc. & $N_{\bar{\nu}_{\mu}}$ after osc. & Statistical \\
        & year$^{-1}$  &   year$^{-1}$     &    year$^{-1}$  & Significance \\
Baseline&  $\exists$ CPT &   $\exists$ CPT   &  $\nexists$ CPT  & $\sigma$ \\
\hline
CERN--SK       &     $1.9$     & $\mathbf{0.9}$   & $\mathbf{1.8\pm1}$  & 0.6  \\
Fermilab--SK    &   $2$      & $\mathbf{1}$       & $\mathbf{1.8\pm1}$  & 0.6  \\
CERN--IceCube  &     $152$     & $\mathbf{76}$    & $\mathbf{152\pm12}$  & $\mathbf{6.1} $  \\
Fermilab--IceCube&   $151$     & $\mathbf{76}$    & $\mathbf{154\pm12}$  &$ \mathbf{6.3}$   \\
\hline
 \textbf{1\% }&   $\langle E_{\mu_{\tau}} \rangle $(Gev) &      $N_{\mu_{\tau}}^{*} $ & $ N_{\bar{\mu}_{\tau}}^{*}$   &  $N_{\bar{\mu}_{\tau}}^{*}$   \\
Baseline&  &   CPT conserved & CPT conserved &  CPT violated\\
\hline
CERN--SK       &     $5.4$     &  0.2 &   0.1     & 0.5 \\
Fermilab--SK    &    $5.6 $      &    0.2 & 0.1     & 0.5 \\
CERN--IceCube  &     $7.3$     &$\mathbf{ 67} $ &  $ \mathbf{33}$    &  $\mathbf{ 16.5 } $   \\
Fermilab--IceCube&   $7.23$      &   $\mathbf{64}$ & $\mathbf{32} $     &   $\mathbf{ 15.3 } $  \\
\hline
\multicolumn{5}{l}{* Considering a $17.4 \% $ branching ratio.}\\
\multicolumn{5}{l}{We remind that the ${\mu_{\tau}}$ energy is nearly $\simeq \frac{1}{3}$ of primary tau.}\\
\multicolumn{5}{l}{ The name ${\mu_{\tau}}$ (in \cite{-7}), has been much later renamed Tautsie-pop in \cite{-8}. }\\

\end{tabular}
\end{small}
\end{center}
\end{table*}

\begin{table*}[htbp]
\begin{center}
\caption{Tau-AntiTau neutrinos in matter by CPT
conserved-violated case.  Estimated total events of charged
current muon neutrino interaction in detector, conversion
probabilities in matter, cross section ratio between
tau-neutrino and muon neutrino, event rates for tau and
anti-tau neutrino (last with both conserved and violated
CPT-symmetry), and noise events by Neutral Current neutrino interactions.}\label{t1a}
\begin{small}
 \tabcolsep 2.1 pt
 \scalebox{0.95}{%
\begin{tabular}{lccccc}
\hline
\hline
 & $N_{\nu_{\mu}}^{CC}$ \tiny no osc.    &   $N_{\bar{\nu}_{\mu}}^{CC}$ \tiny no osc.    & $P_{\nu_{\mu}\rightarrow\nu_{\tau}} $ &  $P_{\bar{\nu}_{\mu}\rightarrow\bar{\nu}_{\tau}}$ \\
Baseline &    in detector  &     in detector   & $\mathbf{\frac{\Delta E}{E} = 10 \%}$  &  $\mathbf{\frac{\Delta E}{E} = 10 \%}$ & $\frac{\sigma_{\nu_{\tau}}}{\sigma_{\nu_{\mu}}} /   \frac{\sigma_{\bar{\nu}_{\tau}}}{\sigma_{\nu_{\mu}}}$ \\
        &       $year^{-1}$&    $year^{-1}$   & $\exists$  CPT    & $\nexists$  CPT &   \\
\hline
CERN--OPERA       & $2847$ &  1423  & $0.015$ & $0.018$     &$0.40\ /\ 0.20$  \\
&&&\\
CERN--SK          & $324$ &    $162$    & $0.962$   & $0.872$    &$0.38\ /\ 0.19$ \\
Fermilab--SK      & $342$ &    $171$    & $0.963$    &  $0.873$    &$0.38\ /\ 0.19$ \\
CERN--IceCube     & $93343$&  $46672$    & $0.945$   &  $0.823$   &$0.41\ /\ 0.20$ \\
Fermilab--IceCube & $93951$&  $46976$  & $0.944$    &  $0.824$    &$0.41\ /\ 0.20$ \\
\hline
&$N_{\nu_{\tau}}^{CC}$ \tiny with osc. & $N_{\bar{\nu}_{\tau}}^{CC}$ \tiny with osc.& $N_{\bar{\nu}_{\tau}}^{CC}$ \tiny with osc.&$ N_{\nu_{i}}^{NC}$ \tiny with osc.&  $ N_{\bar{\nu}_{i}}^{NC}$ \tiny with osc.\\
Baseline & $\exists$  CPT         & $\exists$  CPT     &  $\nexists$  CPT  &  noise $\tau$ like  &  noise $\tau$ like \\
&         $year^{-1}$ &  $year^{-1}$&  $year^{-1}$&  $year^{-1}$&  $year^{-1}$\\
\hline
CERN--OPERA       & $16^{e)}$    & $0.50$     & $0.90$      \\
&&&\\
CERN--SK          & $119$   & $59$    & $54$   & $101$ &   $51 $\\
Fermilab--SK      & $125$   & $63$    & $57$   & $107$ &    $53$  \\
CERN--IceCube     & $36166$ & $18083$ & $15748$ & $29123$ &   $14562 $ \\
Fermilab--IceCube & $36363$ & $18181$ & $15870$ & $29313$ &  $14656$ \\
\hline
&$N_{\nu_{\tau}}^{CC}+N_{\nu_{i}}^{NC} $ & $N_{\bar{\nu}_{\tau}}^{CC}+N_{\bar{\nu}_{i}}^{NC}$ & $N_{\bar{\nu}_{\tau}}^{CC}+N_{\bar{\nu}_{i}}^{NC}$ \\
Baseline & $\exists$  CPT & $\exists$  CPT     &  $\nexists$  CPT   \\
        &  $year^{-1}$    &  $year^{-1}$       &  $year^{-1}$       \\
\hline
CERN--SK          & $220$   & $110$    & $104$   \\
Fermilab--SK      & $232$   & $116$    & $110$     \\
CERN--IceCube     & $65289$ & $32645$ & $30310$  \\
Fermilab--IceCube & $65676$ & $32838$ & $30527$  \\
\hline
\end{tabular}
}
\end{small}
\end{center}
\end{table*}

\begin{table*}
\begin{center}
\caption{Tau-AntiTau neutrinos event rates for reduced experiment, as before, to overall
$10\%$ and to $1\%$.
Statistical significance is referred both for $\nu_{\tau}$, $\bar{\nu}_{\tau}$ only detection, and for CPT cases detection.}\label{t1a}
\begin{small}
\tabcolsep 2.1 pt
 \scalebox{0.99}{%
\begin{tabular}{lccccccc}
\hline
\hline
\textbf{10\%}  &$N_{\nu_{\tau}}^{CC}+N_{\nu_{i}}^{NC} $ &  $\sigma $  & $N_{\bar{\nu}_{\tau}}^{CC}+N_{\bar{\nu}_{i}}^{NC}$ & $\sigma $ &   &$N_{\bar{\nu}_{\tau}}^{CC}+N_{\bar{\nu}_{i}}^{NC}$ \\
Baseline & $\exists$  CPT & for $\nu_{\tau}$ & $\exists$  CPT   & for $\bar{\nu}_{\tau}$ &      & $\nexists$  CPT  & $\sigma $ \\
        &  $year^{-1}$    &           &  $year^{-1}$   &   &  & $year^{-1}$     \\
\hline
CERN--SK          & 22  & 3   &  $\mathbf{11}$    & 2   &  \; \; \;     & $\mathbf{10}$     &   0.2  \\
Fermilab--SK      & 23  & 3   &  $\mathbf{12}$    & 2   &  \ \ \     & $\mathbf{11}$     &   0.2     \\
CERN--IceCube     & 6529 & 45 &  $\mathbf{3264}$  & 32  &   \ \ \    & $\mathbf{3031}$  &   $\mathbf{4.2}$    \\
Fermilab--IceCube & 6568 & 45 &  $\mathbf{3284}$  & 32  &    \ \ \   & $\mathbf{3053}$  &   $\mathbf{4.2}$    \\
\hline
\textbf{1\% }  &$N_{\nu_{\tau}}^{CC}+N_{\nu_{i}}^{NC} $ &  $\sigma $  & $N_{\bar{\nu}_{\tau}}^{CC}+N_{\bar{\nu}_{i}}^{NC}$ & $\sigma $ &   &$N_{\bar{\nu}_{\tau}}^{CC}+N_{\bar{\nu}_{i}}^{NC}$ \\
Baseline & $\exists$  CPT & for $\nu_{\tau}$ & $\exists$  CPT   & for $\bar{\nu}_{\tau}$ &      & $\nexists$  CPT  & $\sigma $ \\
        &  $year^{-1}$    &           &  $year^{-1}$   &   &  & $year^{-1}$     \\
\hline
CERN--SK          & 2   & 0.8 &  1.1                 & 0.6   &  \; \; \;   & 1    &   0 \\
Fermilab--SK      & 2   & 0.8 &  1.2                 & 0.6   &  \ \ \     & 1.1     &   0     \\
CERN--IceCube     & 653 & 14  &  $\mathbf{326}$      & 10  &   \ \ \    & $\mathbf{303}$  &   $\mathbf{1.3}$    \\
Fermilab--IceCube & 657 & 14  &  $\mathbf{328}$      & 10  &    \ \ \   & $\mathbf{305}$  &   $\mathbf{1.3 }$   \\
\hline
\end{tabular}
}
\end{small}
\end{center}
\end{table*}



\clearpage

\section{Conclusions and discussion}
Here we offered the estimate of the neutrino event rate in different scenario (CERN or FNAL beaming $\nu$ to SK or ICECUBE),
at different approximation (mostly keeping care of the matter influence), at different experiment set up (like OPERA, or just a fraction $10\%$ or $1\%$ percent of it) with energy dispersion $\frac{\Delta E}{E}$ as large as $10\%$, $20\%$. We estimated the $\nu_{\mu}$, $\nu_{\tau}$  into muon or tau (and the same for the anti particles) event rate at each configuration finding high rate to detect $\tau$, $\bar{\tau}$  appearance; we were inspired to disentangle any hypothetical $\nu_{\mu}$, $\bar{\nu_{\mu}}$ CPT violation  even at level below MINOS detection threshold.
We noted how the neutrino mixing under present CPT violated presence may play a minor  role in atmospheric neutrino events in Deep Core (mostly at 3-5 channels): it is difficult to use the $\nu_{\mu}$ arrival direction because poor angular resolution, to estimate the oscillation parameters.  Electron neutrinos, neutral current events are polluting any atmospheric tau appearance. Therefore we  consider an artificial scenario (a new OPERA like experiment), where we know a priori the source $\nu_{\mu}$ energy (with some accuracy) and the source-detector distance. This scenario cut at once atmospheric uncertainty and the additional noises due to $\nu_e$, $\bar{\nu}_e$ presence, and the $\nu_{\mu}$, $\bar{\nu}_{\mu}$ different arrival distances and energies. These longest baseline neutrino experiment are not much different from already proposed ones \cite{13} (for hundred kiloton or SuperK), aimed to test the MSW resonant effect. But at the energies here considered and the largest distances and widest detector masses lead to largest oscillation phase and the an optimal  $\nu_{\tau}$ birth and detection. As shown in tables 8,9 we expect for CERN - IceCube-DeepCore scenario, even a $1\%$ of OPERA efficiency, a rate of $652$ tau like events ($291$ noise + $361$ signal) a year, offering  nearly $\geq 14 \sigma $ in one year of experiment for tau neutrino appearance. A comparable solid result  nearly $\geq 10 \sigma $ will be tested for anti-tau appearance. These $\tau$ like events will be rising as light explosion observable within a quadruplet  clustering (triplet or quintuplet may also appear) in Deep Core DOMs.

At the same time in the same scenario sending $\nu_{\mu}$ we expect (because energy indeterminacy) the appearance of nearly $152$ $\mu^-$ events even at worst scenario (a minimal set up and maximal $\frac{\Delta E}{E}$ as large as $20\%$). Their negligible number differ from no-mixing case (nearly expected $933$ $\mu^-$ events) offering a decisive correlated tau appearance coeval imprint and a solid muon flavor disappearance signature. It is important to note that in DeepCore a $\nu_{\mu}$ will release an average $52\% $ of its energy $(\langle E \rangle \simeq 0.52 E_{\nu} \simeq 11 GeV)$. The average muon length projected within $cos(\theta_{Cern - IceCube}) \simeq cos(22°)\simeq 0.92$ will light nearly 7 optical phototube at once leading to  characteristic 652  quadruplet versus 152   event rate a year. We remind that the neutrino beaming bunches from CERN of from FNAL are born in very narrow windows in Universal Time, leading to an integral time, each a year, within or below ten seconds. Inside this time lapse one would expect a tiny fraction ($\sim 3 \cdot 10^{-7}$)of atmospheric $\nu_{\mu}$ (or $\nu^{NC}$) event, just one atmospheric event every century. Therefore the neutrino beaming is completely noise free from cosmic events. On the other side the beaming of anti muon primary (as shown in table 4-5-6-7) leading to $\bar{\nu}_{\mu}$ and $\bar{\nu}_{\tau}$ in either CPT conserved and violated scenario, makes very different signals: $326\pm 18$ conserved versus $303\pm 15.6$ violated cases: a barely negligible signal.

Moreover, at $1\%$ of OPERA efficiency, while sending $\bar{\nu}_{\mu}$, we will observe in the same detector different signature in $\mu^+$ for either suppressed oscillations, or for the CPT violating different mixing parameters: 76 events (CPT conserved with energy spread $\frac{\Delta E}{E}$ as large as $20\%$) versus $152\pm 12.2$ (CPT violated), leading to at least $6 \sigma$ significance to disentangle CPT violation puzzle. The signal is expected to be loud and clear: indeed, we do foresee quadruplets for the $\tau$ like lights  and the $E_{\mu^+} \simeq 0.66  E_{\bar{\nu}_{\mu}}$ will lead to quite energetic and larger $\mu^+$ tracks, corresponding (in present Cern-ICECUBE geometry) to 9 photomultipliers (DOM) channel events, on average. The comparison between CPT conserved events (76 $\mu^+$ in 9 channel) and CPT violated ones (152 $\mu^+$ in 9 channel) is a  statistical sharp mark.
Let us remind that the whole IceCube contain an order of magnitude more photomultipliers (nine times DeepCore) and it might be an additional detector even with a partial track trigger. It should be also remind the extreme accuracy in testing muon disappearance (152 event a year by noise and energy  dispersion) respect to no oscillation case (933 events a year), that may provide a crucial test for a tuned $\sin(\theta_{23})\simeq 0$  opening to new windows to eventual hidden lepton symmetry. Also in the anti-neutrino muon version (($\sim 76$ event a year by noise and energy  dispersion) respect to no oscillation case ($\simeq 467$ events a year)). The higher dense PINGU array in a near future may also dig into the lower energy neutrino range testing the $\theta_{13}$ angle value as well as the eventual inverted hierarchy neutrino mass (See Fig. \ref{Theta-_13}, \ref{IHmu}). Beaming neutrino along the whole Earth may become a day, a three and a half times "faster than light" $\Delta T \simeq \frac{ R_{\bigoplus} \cdot (\pi - 2)}{c}$  telegraph able to fast message within a second fraction along continents any  trade  message across the Earth. In CERN-Deep Core configuration the neutrino reaches the detector with a precursor time of just $1.03 \cdot 10^{-2} s$. Such a large distances and time flight $3.93 \cdot 10^{-2}$ s offer the possibility for  a sharper neutrino  speed test (a questionable on fashion experiment in our days \cite{-20}).
Different applications as the economic ones may be also of great interest. In the same experiment we are considering, it is  possible to test at 6 GeV deviations due to different $\theta_{13}$ mixing angles (see Fig.\ref{Theta-_13}). In the low energy ranges within our proposed experiment it is possible to disentangle the thin neutrino mass splitting as shown in figure \ref{IHmu},\ref{IHe}, in different hierarchy neutrino mass models.  It may sound exaggerated   that the largest experiment on Earth is needed by  beaming to longest distance,  weakest $\nu_{\mu}$ particles toward largest existing detector, in order
to disentangle their lightest lepton mass splitting, their  angles, and any eventual hidden CPT secrets. For these reasons such an experiment at $1\%$ OPERA size seem  to us very affordable  and attractive: one $\tau$ a day (or one  anti  $\tau$ every two days) versus one $\tau$  a year (actually in few years) in OPERA is indeed amazing, also in view of nearly $6 \sigma$ test for any hypothetical CPT violation  within earliest (2010) CPT deviations and present (2012) MINOS bounds.

\section{Acknowledgments}
We are grateful to Prof. B. Mele, P. Lipari, Dr. P. Oliva and in particular  L. Ludovici  for very useful discussions, comments and suggestions.
\section{Dedication to Nicola Cabibbo}
This paper begun soon after the MINOS first result on middle  June 2010: we had discussed, by phone, the early backbone of our paper with
 Prof. Nicola Cabibbo, who was already in disease. He did appreciated the proposal and he did offer  attention and discussions on our preliminary elaborations. Indeed his support greatly encouraged  our effort, making  somehow himself a promoter or even a probable  co-author of the present article. Sadly and suddenly Prof. N. Cabibbo disappeared just on $16$ August 2010. We  wish at least to dedicate this paper in his memory. He  was the founder of the main quark mixing operator (known today as the Cabibbo Kobayashi Maskawa matrix), a basic discovery  deeply connected to the coeval lepton neutrino matrix (the Pontecorvo Maki Nakagawa Sakata matrix), whose role is the leading one to the complex muon-tau metamorphosis  in flight, that is the road map of our article. Nicola Cabibbo has been  always extremely friendly (incidentally  Habbib in Hebrew means friend), sharp  in his passion and love for the truth. His interests were not  on elementary particle physics alone, but they extended in a broadest Science fields. For instance he did support the research (offering also a   memory) of Prof. A. Cacciani, a rarest expert on Solar physics lost four years ago; he did also support once youngest researchers (now, the oldest of us)  in astro-particle. Finally he dedicate last few months and weeks in unique effort to elaborated photo masterpiece composition. He was surprising us along his  scientific life with wisdom, synthesis and insight.
 At his end, in his suffering late times, he have shown his greatness and kindness by unexpected explosive texture of forms and colors.  Possibly he is finally enjoying of rest with B. Pontecorvo, W. Pauli and E. Fermi from above, in deepest contemplation of science  harmony and secrets, while smiling on our human smallness. Grazie Nicola.

\begin{figure}[h!]
\begin{center}
\includegraphics[angle=0,scale=0.23]{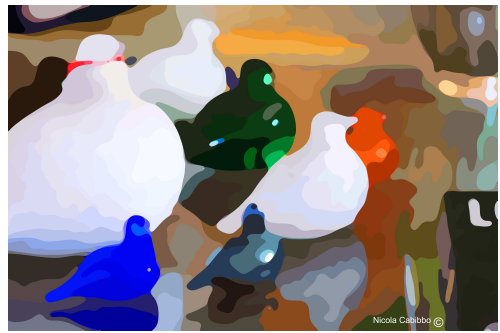}
\caption{ One of the amazing  Cabibbo photo-color composition, in a bath of colors, made during last months of life. }\label{cabibbo}
\end{center}
\end{figure}

\begin{appendix}
\section{ A-The general  Pontecorvo-Maki-Nakagawa-Sakata mixing mass matrix}

In particle physics, the Pontecorvo-Maki-Nakagawa-Sakata matrix (PMNS matrix),
 neutrino mixing matrix, is a generalization of the Pontecorvo two flavor case.
 This generalization is the analogous one of the Cabibbo-Kobayashi-Maskawa quark mixing, or CKM matrix. It is a unitary matrix which contains
 information on the mismatch of quantum states of leptons when they propagate freely and when they take part in the weak interactions.

Let us remind the main matrix that we shall consider in neutrino mixing:

$$
U\,=\,\left(\begin{array}{ccc}
U_{e1} & U_{e2} & U_{e3} \\
U_{\mu1} & U_{\mu2} & U_{\mu3} \\
U_{\tau1} & U_{\tau2} & U_{\tau3} \\
\end{array}\right)\,=
$$

$$
=\,\left(\begin{array}{ccc}
1 & 0 & 0 \\
0 & c_{23} & s_{23} \\
0 & -s_{23} & c_{23} \\
\end{array}\right)\cdot
\left(\begin{array}{ccc}
c_{13} & 0 & s_{13}e^{-i\delta} \\
0 & 1 & 0 \\
-s_{13}e^{i\delta} & 0 & c_{13} \\
\end{array}\right)
\cdot\left(\begin{array}{ccc}
c_{12} & s_{12} & 0 \\
-s_{12} & c_{12} & 0 \\
0 & 0 & 1 \\
\end{array}\right)\cdot
\left(\begin{array}{ccc}
e^{i\alpha_{1}/2} & 0 & 0 \\
0 & e^{i\alpha_{2}/2} & 0 \\
0 & 0 & 1 \\
\end{array}\right)\,=
$$

\begin{equation}
=\,\left(\begin{array}{ccc}
c_{12}c_{13} & s_{12}c_{13} & s_{13}e^{-i\delta} \\
-s_{12}c_{23}-c_{12}s_{23}s_{13}e^{i\delta} & c_{12}c_{23}-s_{12}s_{23}s_{13}e^{i\delta} & s_{23}c_{13} \\
s_{12}s_{23}-c_{12}c_{23}s_{13}e^{i\delta} & -c_{12}s_{23}-s_{12}c_{23}s_{13}e^{i\delta} & c_{23}c_{13} \\
\end{array}\right)
\cdot\left(\begin{array}{ccc}
e^{i\alpha_{1}/2} & 0 & 0 \\
0 & e^{i\alpha_{2}/2} & 0 \\
0 & 0 & 1 \\
\end{array}\right)
\end{equation}

 Where $s_{ij}=\sin\theta_{ij}$ and $c_{ij}=\cos\theta_{ij}$ with $i,j\,=\,1,2,3$. The first matrix is related to atmospheric mixing, the second one is an unknown composition of flavors also related to CP violated terms,the third matrix is correlated
to solar neutrino mixing. \\The present values for angles are: \\
 $\sin^2(2 \theta_{13})=0.092$ or
 \(\theta _{13}\)= $10.3^{\circ}$ , \\
$\sin^2(2 \theta_{23})=0.995$ or
  \(\theta _{23}\)= $42.97^{\circ}$,\\
$\sin^2(2 \theta_{12})=0.86$ or
 \(\theta _{12}\)= $34.01^{\circ}$ ;\\
 we assume for sake of  simplicity $\alpha_{1} = \alpha_{2} = 0 $.

The mixing matrix U is thus responsible for rotation between flavour and mass eigenstates:

\[\left.\left|\nu _{\alpha }\right.\right\rangle =\sum _k U_{\text{$\alpha $k}}^* |\nu _k\rangle  ,\text{     }(\alpha = e,\mu ,\tau \text{ $\, $}\,
\, \, \text{and}\, \, \, \,  k=1,2,3)\]

Resolving the Schrodinger equation for mass eigenstates time evolution leads to the following transition amplitude :

\[A_{\nu _{\alpha }\rightarrow \nu _{\beta }}(t)=\sum _k U_{\text{$\alpha $k}}^*U_{\text{$\beta $k}}\text{ $\, $}e^{-i E_kt}\]

Since we are dealing with relativistic neutrinos, the following approximations are valid:

\[E_k\simeq E+\frac{m_k^2}{2 E},\text{  }t=L\]

and thus the transition probability is:

\begin{equation}
P_{\nu _{\alpha }\rightarrow \nu _{\beta }}(L,E)=\sum _{k,l} U_{\text{$\alpha $k}}^*U_{\text{$\beta $k}} U_{\text{$\alpha $l}} U_{\text{$\beta
$l}}^*\, \exp \left(-i \frac{\text{$\Delta $m}_{\text{kl}}^2L}{2 E}\right)
\end{equation}

\begin{equation}
=\sum _{k,l} U_{\text{$\alpha $k}}^*U_{\text{$\beta $k}} U_{\text{$\alpha $l}}U_{\text{$\beta $l}}^*\, \exp \left(-i\ 2.54\ \frac{\text{$\Delta $m}_{\text{kl}}^2}{{eV}^2} \frac{{GeV}}{\text{E}} \frac{{\text{L}}}{km} \right)
\label{equation1}
\end{equation}

and for the antineutrinos:

\begin{equation}
P_{\bar{\nu}_{\alpha }\rightarrow \bar{\nu}_{\beta }}(L,E)=\sum _{k,l} U_{\text{$\alpha $k}}U_{\text{$\beta $k}}^* U_{\text{$\alpha $l}}^* U_{\text{$\beta $l}}\, \exp \left(-i\ 2.54\ \frac{\text{$\Delta $m}_{\text{kl}}^2}{{eV}^2} \frac{{GeV}}{\text{E}} \frac{{\text{L}}}{km} \right)
\label{equation1}
\end{equation}

where \(\text{$\Delta $m}_{\text{kl}}^2=m_k^2-m_l^2\).

Usually the two flavour mixing case is considered:
\[P_{\nu _{\alpha }\rightarrow \nu _{\beta }}(L,E)=\sin ^2(2\theta )\sin ^2\left(1.267\frac{\text{$\Delta $m}^2}{{eV}^2} \frac{{GeV}}{\text{E}} \frac{{\text{L}}}{km}\right)\]

which is justified in the assumption that the $\theta_{13}$ mixing angle is small, and  $\Delta m_{23} \ll \Delta m_{12}$. A more accurate formula for three
flavour can be used, for instance $P_{\nu _{\mu }\rightarrow \nu _{\tau }} $:

\[P_{\nu _{\mu }\rightarrow \nu _{\tau }}(L,E)=\sin ^2\left(2\theta _{23}\right)\cos ^4\left(\theta _{13}\right)\sin ^2\left(1.267 \frac{\text{$\Delta $m}_{\text{23}}^2}{{eV}^2} \frac{{GeV}}{\text{E}} \frac{{\text{L}}}{km}\right)\]

These are useful analytical approximation, however in view of a more precise probability calculation we performed a three flavour numerical calculation using equation \ref{equation1}.
The deviation between two and three components by analytical equation in vacuum is small and fall within energy and length uncertainty, but in matter it's worth to perform numerical calculation since analytical solution of Schroedinger equation with matter potential in three flavor case is a more complex task. In general three-flavour treatment can add corrections up to $\sim 10 \%$ respect to two-flavour one \cite{14}.\\

The evolution equation for three neutrino mixing in matter reads :
\begin{equation}
i\frac{d}{d x}\Psi _{\alpha }=H \Psi _{\alpha } ,\ \  \text{with} \ \  H =\frac{1}{2 E}\left(U M U^{\dagger }+A\right),  \\\ \alpha =1,2,3
\end{equation}

which is a Schrodinger equation with a matter potential \(A\),

\[A=\left(
\begin{array}{ccc}
 A_{\text{CC}} & 0 & 0 \\
 0 & 0 & 0 \\
 0 & 0 & 0
\end{array}
\right),\text{ $\, \, \, $ }M=\left(
\begin{array}{ccc}
 0 & 0 & 0 \\
 0 & \text{$\Delta $m}_{21}^2 & 0 \\
 0 & 0 & \text{$\Delta $m}_{31}^2
\end{array}
\right),\text{ $\, \, \, $}\Psi _{\alpha }=\left(
\begin{array}{c}
 \psi _{\text{$\alpha $e}} \\
 \psi _{\alpha \mu } \\
 \psi _{\alpha \tau }
\end{array}
\right)\]

where \ $A_{CC}= 2 \sqrt{2}  E G_F N_e=0.76 \, \cdot 10^{-4} eV^2  ( \frac{E}{GeV}) (\frac{\rho}{g/cm^{3}}) $ is the potential only due to \(\nu _e\) charged current interaction.

For the antineutrino interaction in matter, the potential $A_{CC}$ changes sign. This introduces a deviation in oscillation probability due only to matter effects for antineutrino respect neutrino oscillation, which is nevertheless negligible at energies above $\sim 5-6$ GeV, and anyway nearly undetectable with present
detectors energy resolution. Therefore the only difference between neutrino and antineutrino oscillation probability is eventually due only to CPT violating parameters.

This equation is a set of 9 differential equations whose solutions we can call $P_{\alpha \beta}= \Psi_{\alpha \beta}^{2}$, with $\alpha, \beta =e,\mu, \tau$ (following \cite{13},\cite{14}) resolved numerically in
 the exact earth density profile (which is different for the two different baselines, respectively to SuperKamiokande and to IceCube), with boundaries conditions $P_{\alpha \beta}(0)$ equal to 1 for the starting flavour, 0 for the others.
 The results are shown in figures 5. It is clearly visible the small difference from oscillation in vacuum, at energies considered ($\sim $ 20 GeV), however it substantially deviates from the vacuum case at lower energy, because approaching the resonance energy in matter (MSW effect). The numerical solutions have the following form, for instance: \\
 $P_{\mu \tau}= \left|-0.5 e^{-0.000019 i x}+0.31 e^{-0.00074 i x}+0.19 e^{-0.000986 i x}\right|^2 $ \\
at 5 GeV in matter density 4.5 $g/cm^{3}$. Numerical solutions in exact earth matter profile have more complex functional form.

\begin{figure}
\begin{center}
\includegraphics[angle=0,scale=.28]{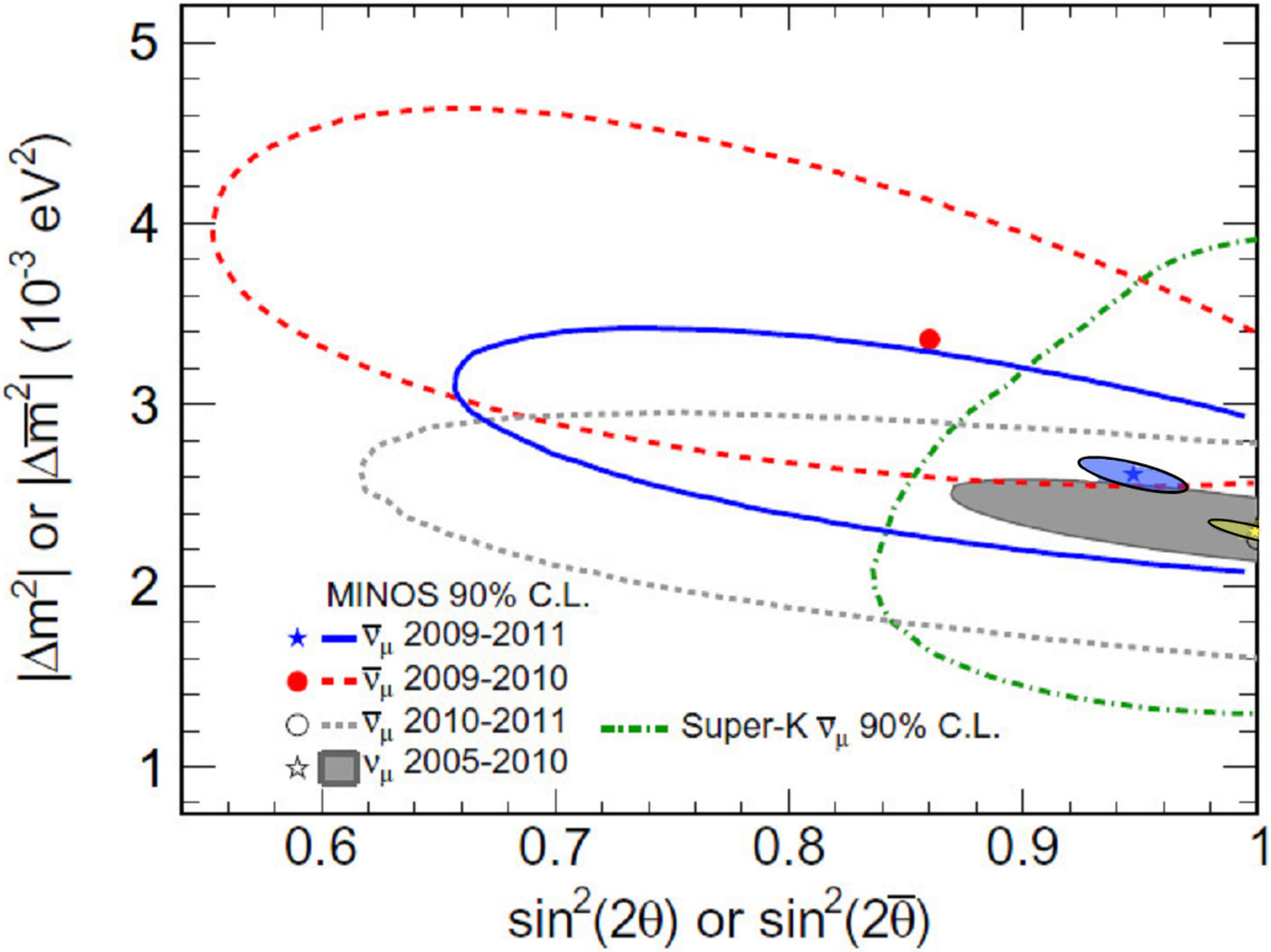}
\caption{ Neutrino and anti-neutrino muon  oscillation probability in $\sin(2\cdot \theta_{23})=1$ and  $\Delta m^{2}=(2.35+0.11-0.08)\cdot 10^{-3}eV$ as in old data by MINOS and SK. Also the old
 anti-neutrino muon  oscillation probability  (MINOS, Neutrino 2010 Conference, $14-June-2010$) into anti-tau neutrino CPT violated parameters was
   $\Delta \overline{m}_{23}^{2} =(3.36^{+0.45}_{-0.40}($stat.$)\,\pm\,0.06($syst.$))\cdot 10^{-3}$ eV$^{2}$ and $\sin^{2}(2\overline{\theta}_{23})\,=\,0.86\,\pm\,0.11($stat.$)\,\pm\,0.01($syst.$)$.
   On the contrary, the recent parameters are more comparable with the CPT conserved ones: $\Delta \overline{m}_{23}^{2} =(2.62^{+0.31}_{-0.28}($stat.$)\,\pm\,0.09($syst.$))\cdot 10^{-3}$ eV$^{2}$, $\sin^{2}(2\overline{\theta}_{23})\,=\,0.945$.
  The early MINOS discordance was about $2.5$ sigma, but the most recent one is  within one sigma consistent with the CPT conserved case. Our beaming across the Earth might reach a discrimination described somehow by the inner smaller ellipses, whose extension at 6 sigma may disentangle even last eventual MINOS tiny CPT discordance. }\label{MINOS-ellipse}
\end{center}
\end{figure}
\begin{figure}
\begin{center}
\includegraphics[angle=0,scale=.28]{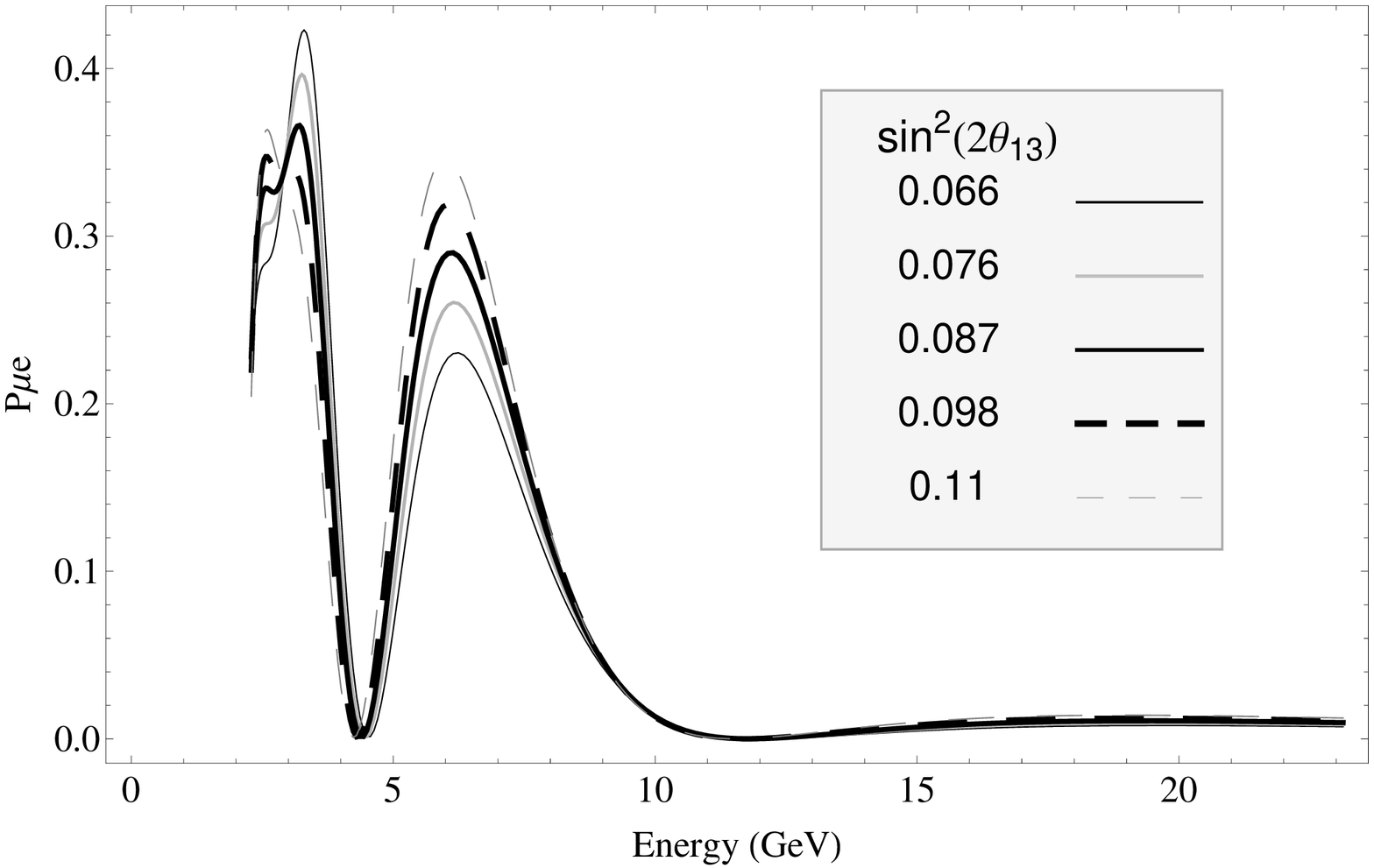}
\caption{ The conversion probability of muon neutrino into electron, for different $\sin(2 \theta_{13})^{2}$ values,  a probability deviations that may test at low energy the disappearance of muons tracks and the appearance of electron showers (similar to tau ones). The present Deep Core  array can hardly be able to reveal such a small energy signals  (muon tracks versus electron showers), while future more dense PINGU array  may be a better tuned detector.
}\label{Theta-_13}
\includegraphics[angle=0,scale=.28]{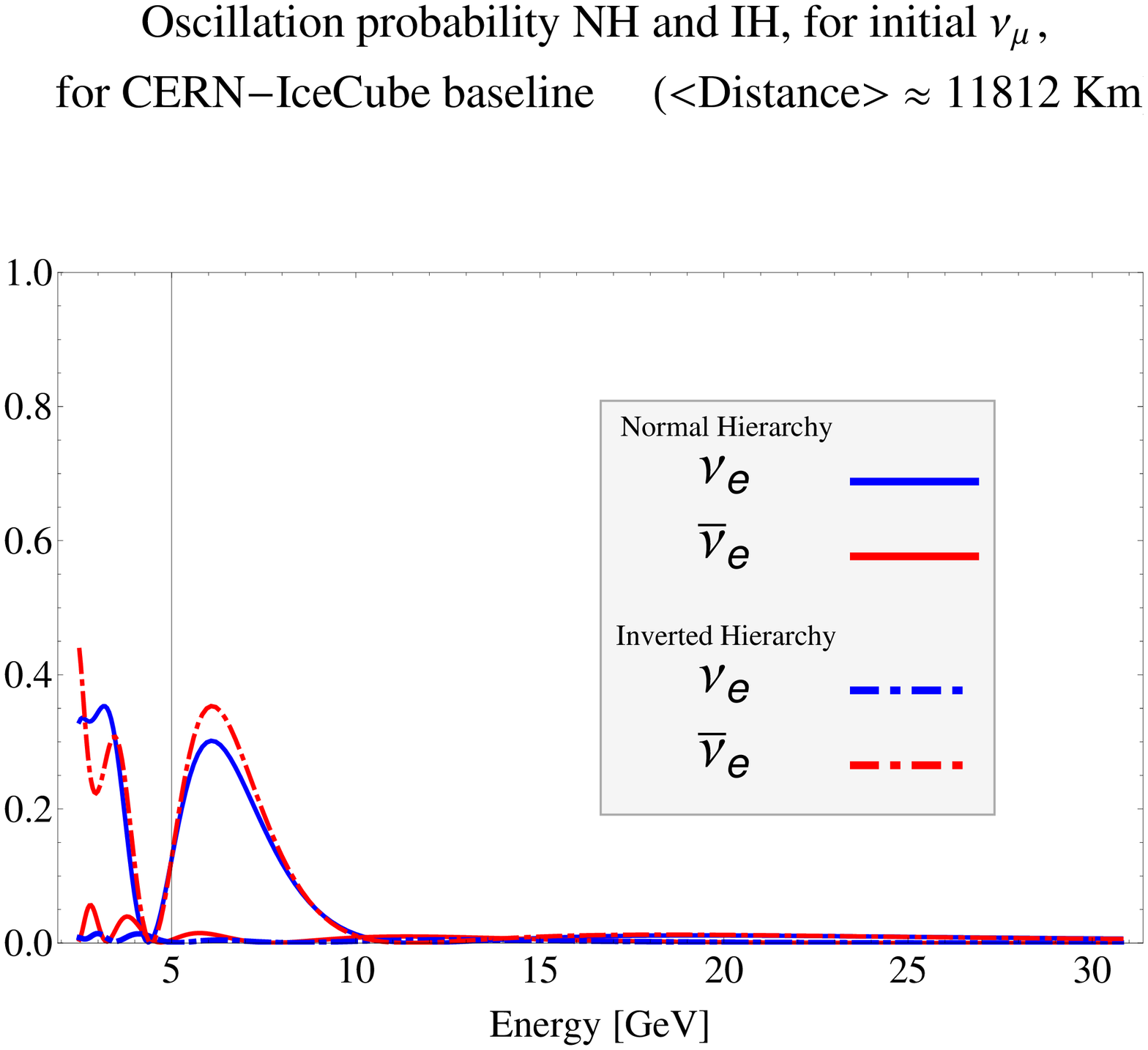}
\caption{The conversion probability of muon neutrino into electron,in normal or inverted neutrino mass hierarchy. As shown in the figure there is a remarkable deviation from normal neutrino mass hierarchy and inverted one mostly in the low energy region ($\simeq 6$ GeV), where  the $\bar{\nu}_{e}$ appears (in the inverted case) at a $38 \%$ probability rate (born by a $\bar{\nu}_{\mu}$ conversion), while it is nearly absent $\simeq 2 \%$ probability rate in normal neutrino mass hierarchy.}\label{IHmu}
\includegraphics[angle=0,scale=.28]{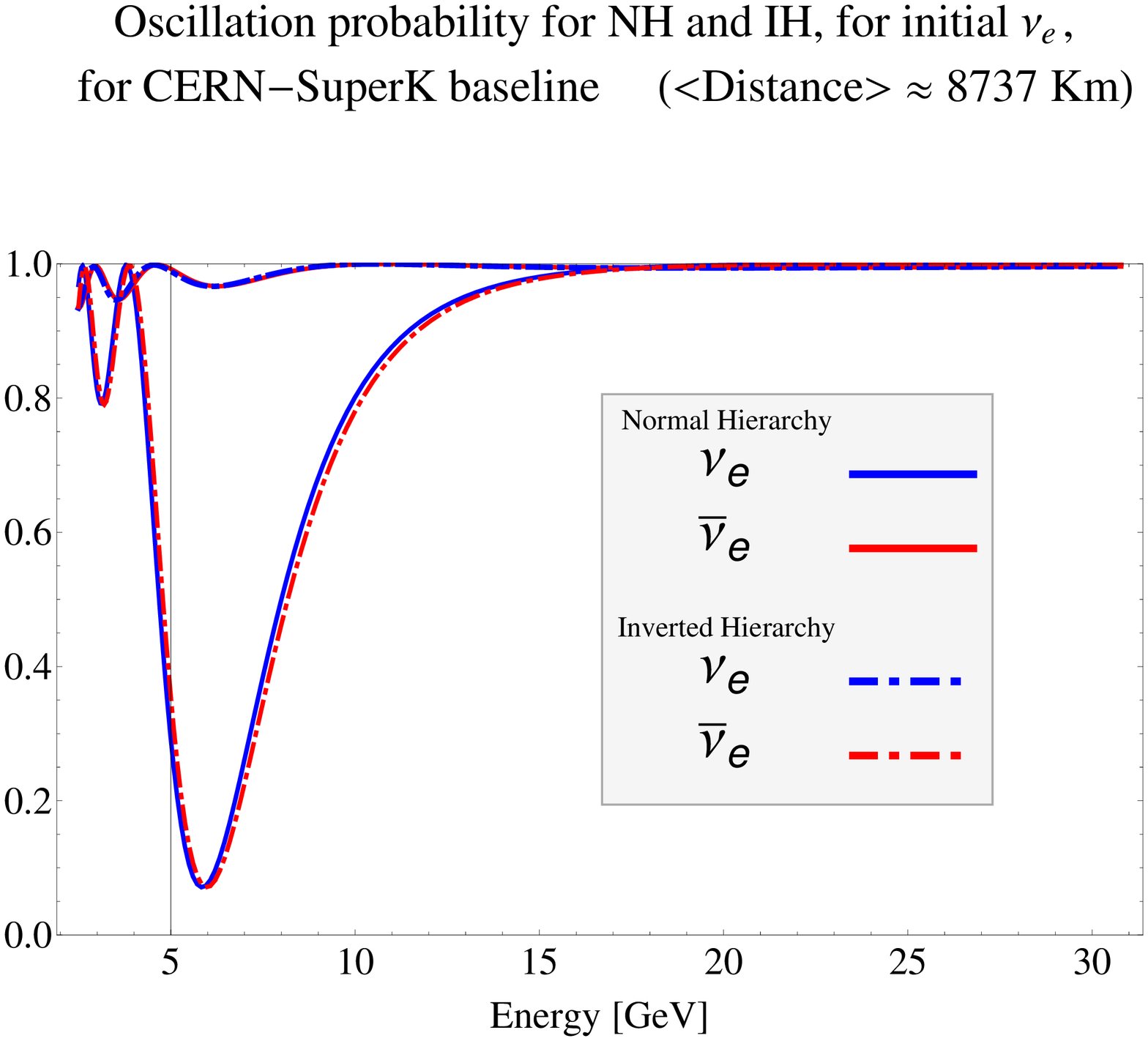}
\includegraphics[angle=0,scale=.28]{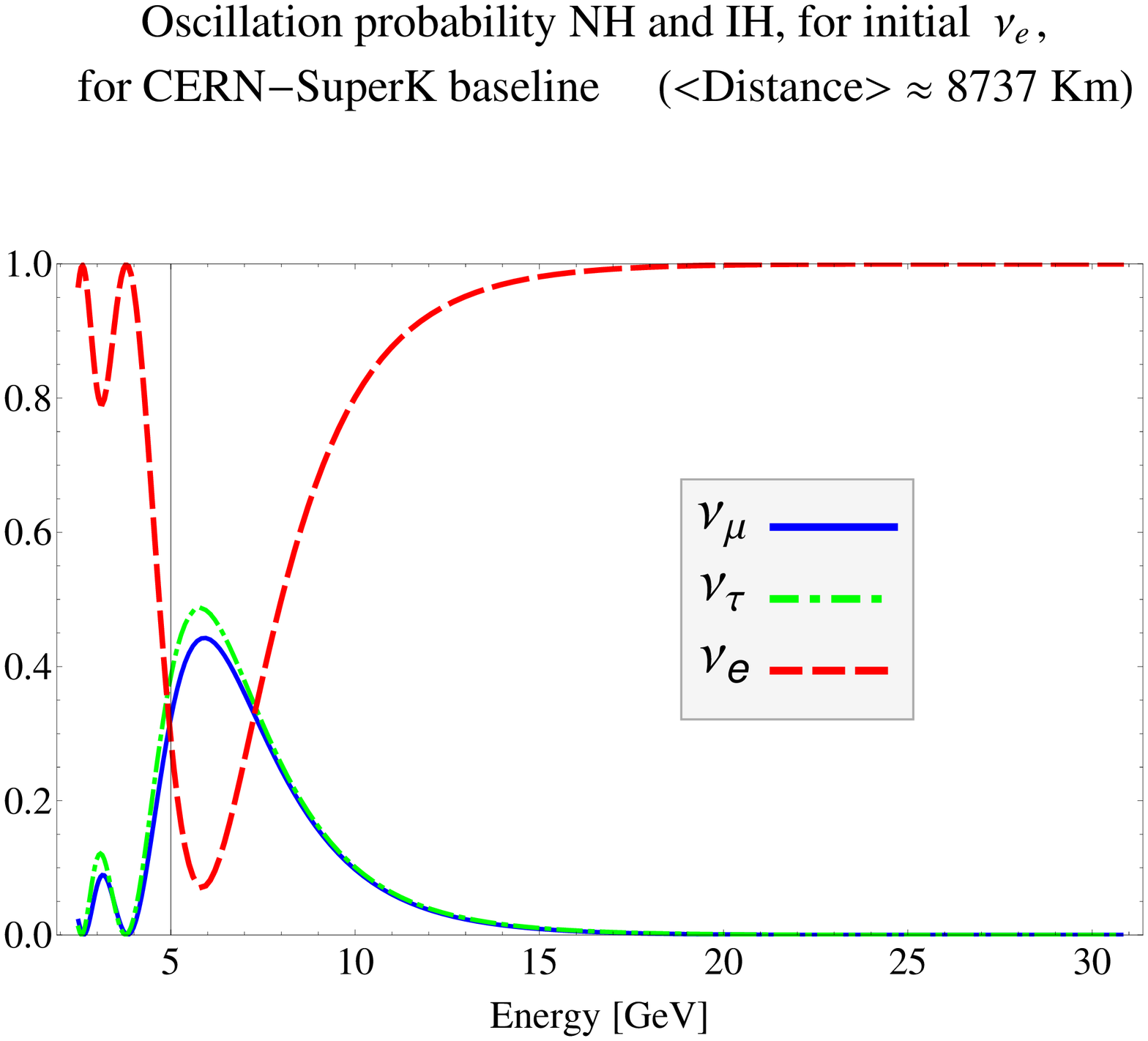}
\caption{Left: the electron neutrino survival probability $P_{\nu_e\rightarrow \nu_e}$ and $P_{\bar{\nu}_e \rightarrow \bar{\nu}_e}$, in normal or inverted neutrino mass hierarchy. . As shown in the figure there is a remarkable deviation from normal neutrino mass hierarchy and inverted one mostly in the low energy region ($\simeq 6$ GeV), which turns to be higher in the SuperK baseline respect the IceCube one. A $\nu_e , \bar{\nu}_e$ beam is suitable for best hierarchy model discrimination.
Right: the oscillation probability $P_{\nu_e\rightarrow \nu_{\alpha}}$ for the SuperK baseline show that there is non-negligible conversion
into muon neutrino, possibly detectable in future $\nu_e$ beam.}\label{IHe}
\end{center}
\end{figure}


\begin{figure}[h]
\begin{center}
\includegraphics[angle=0,scale=.12]{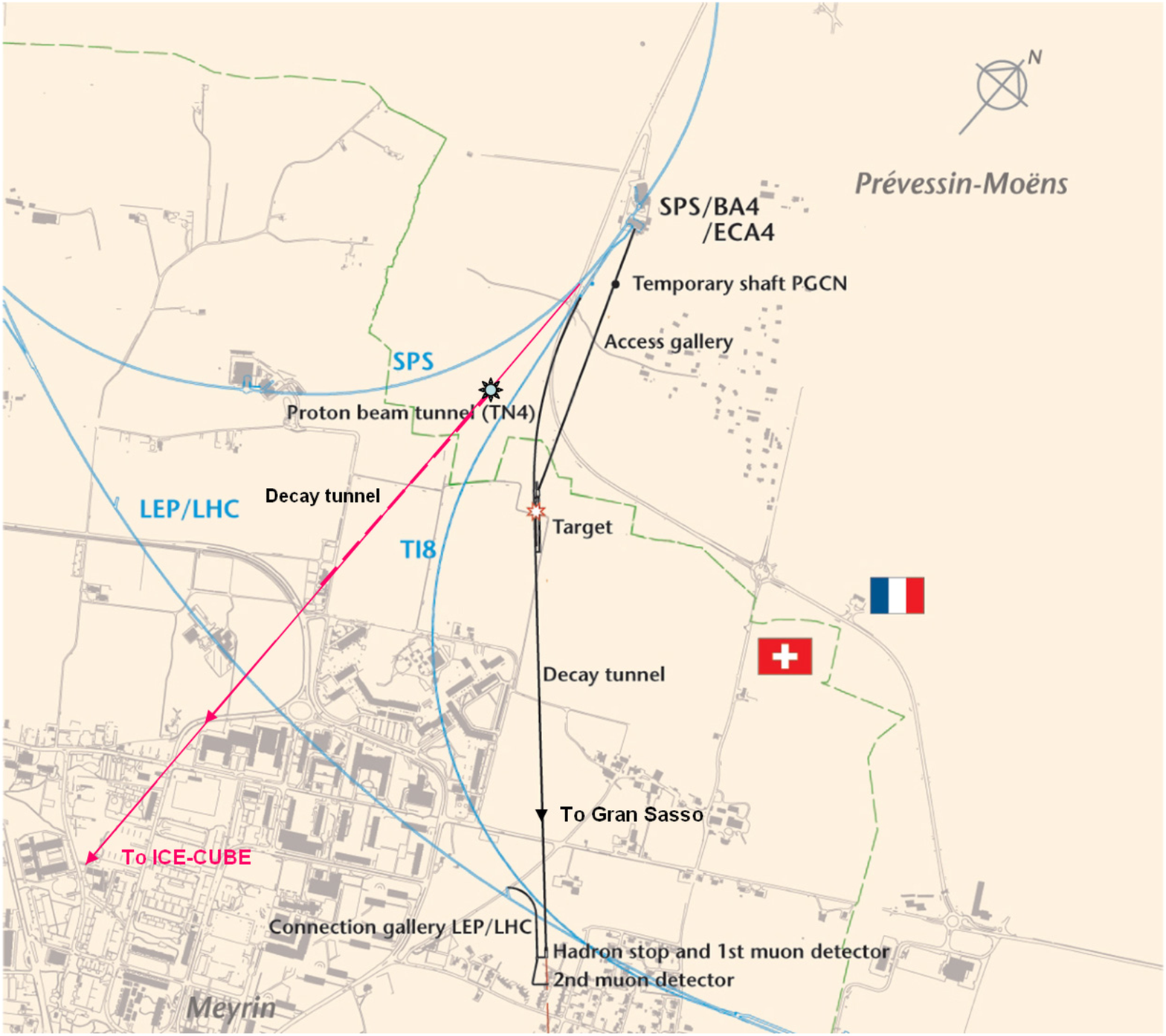}
\includegraphics[angle=0,scale=.15]{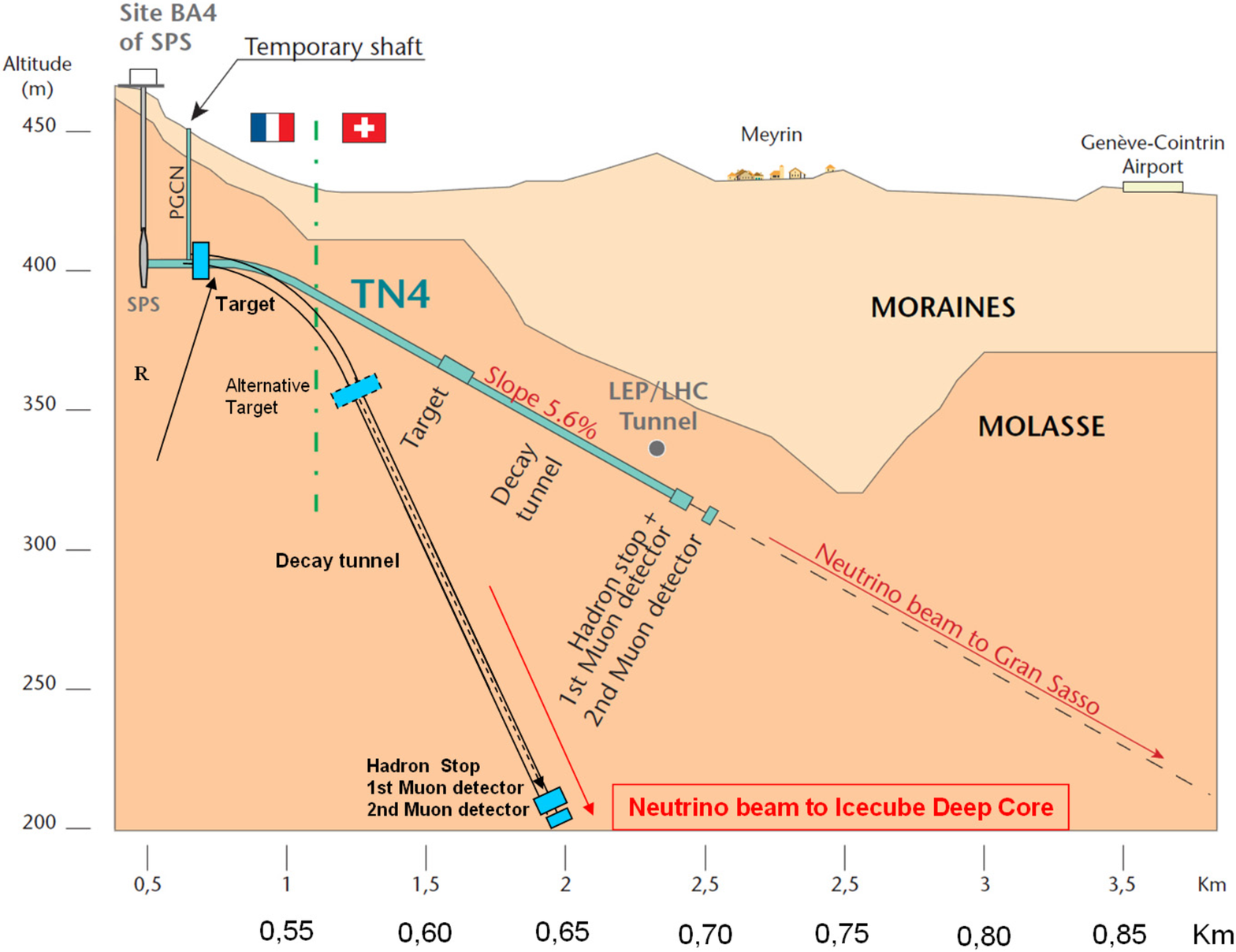}
\caption{Beaming from CERN   at $22$ GeV neutrino to IceCube. The geometry of our additional image to IceCube (to OPERA) is well in proportion and
it is described by a lower scale below. The curvature radius of the tunnel will be $171$ m. for a bending magnetic field at nominal 1 Tesla,
 turning down by $67.82^{o}$ toward South  Pole. The total arc length is $202$ m. long while the height below (by deflection) the level is first $106$ m; the additional decay tunnel is suggested $92$m , leading to only
 $ 5 \%$ of the  distance for decaying pions (with respect to OPERA). This bending and distance might naturally filter as a spectrum-meter, positive (or negative) pion and muon neutrino secondaries. This is the most economic scenario at $1\%$ discussed in tables. }\label{Tunnel-CERN}

\vspace{2cm}

\includegraphics[angle=0,scale=.13]{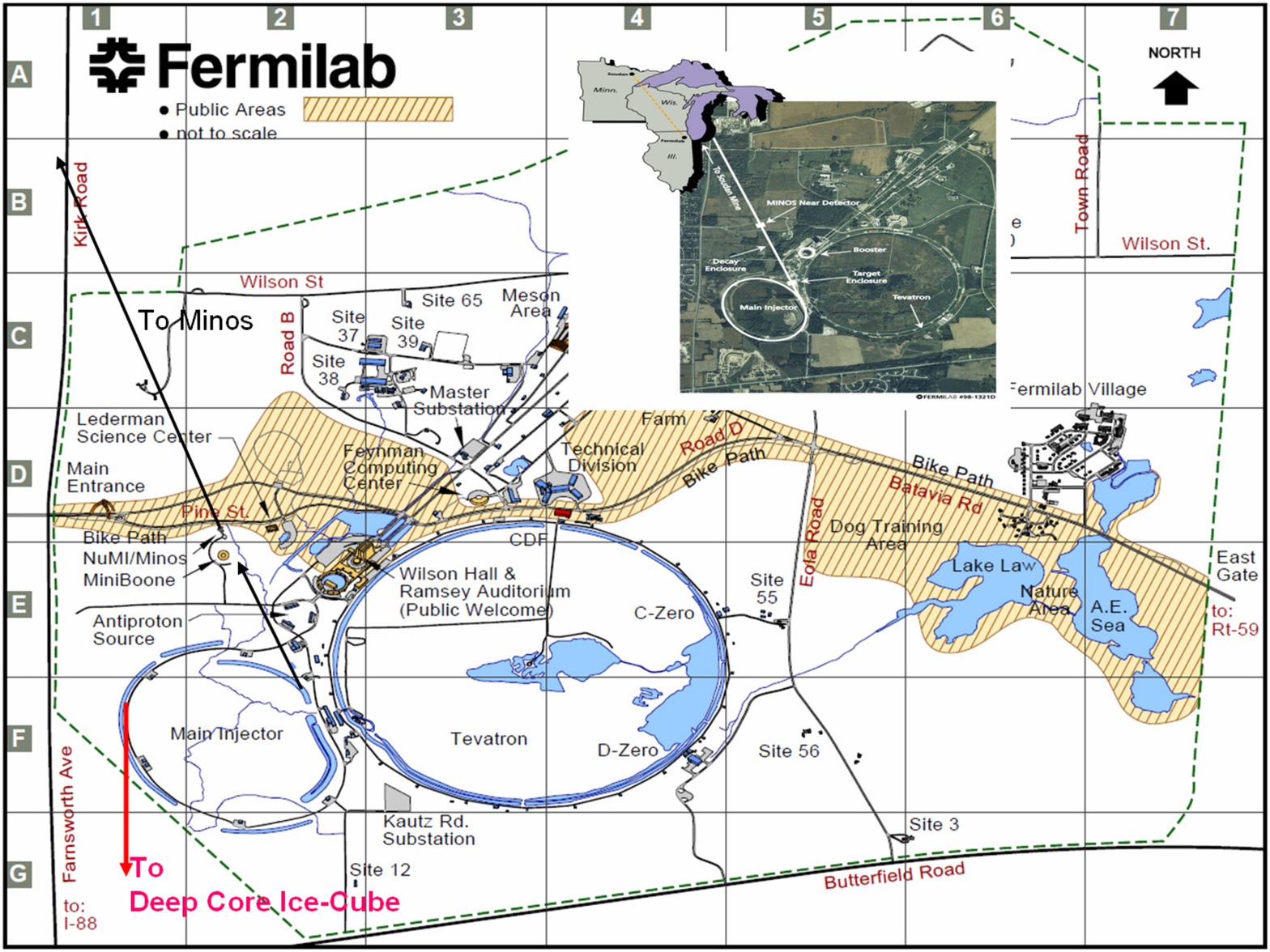}
\includegraphics[angle=0,scale=.13]{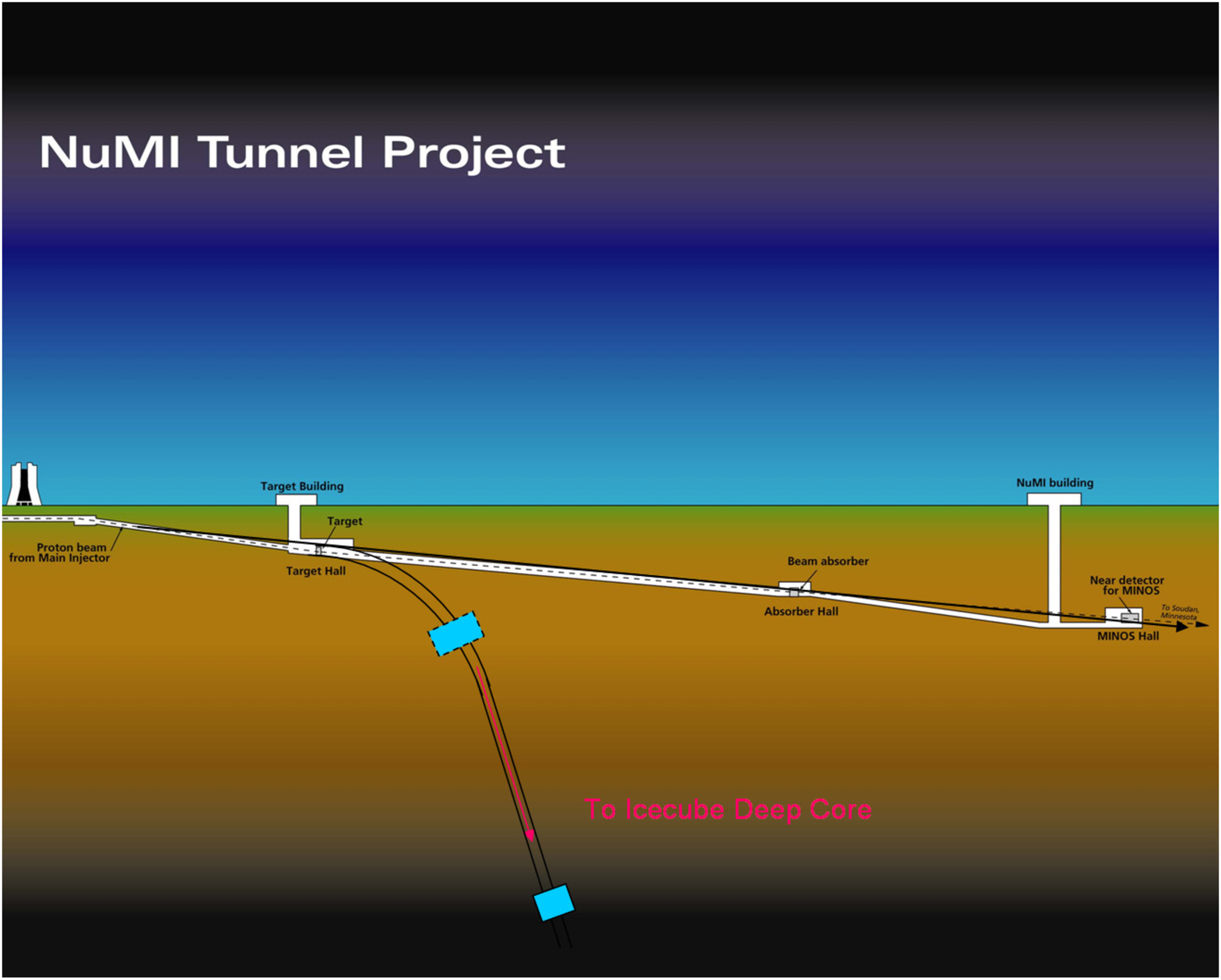}
\caption{Beaming from  FNAL a $21$ GeV neutrino to IceCube. The bending geometry of our additional image to IceCube  is in proportion.
The curvature radius of the tunnel will be $169$ m. for a bending magnetic field at nominal 1 Tesla,
 turning down by $65.67^{o}$ toward South  Pole. The total arc length is $194$ m. long while the height below (by deflection) the level is first $99$ m; the additional decay tunnel is suggested $90$m , leading to only  $ 5 \%$ of the distance for decaying pions (with respect to OPERA). This bending and distance might naturally filter as a spectrometer, positive (or negative) pion and therefore their muon neutrino secondaries.This is the most economic scenario at $1\%$ discussed in tables.}\label{Tunnel-Fermi}
\end{center}
\end{figure}

\begin{figure}[h]
\begin{center}
\includegraphics[angle=0,scale=.17]{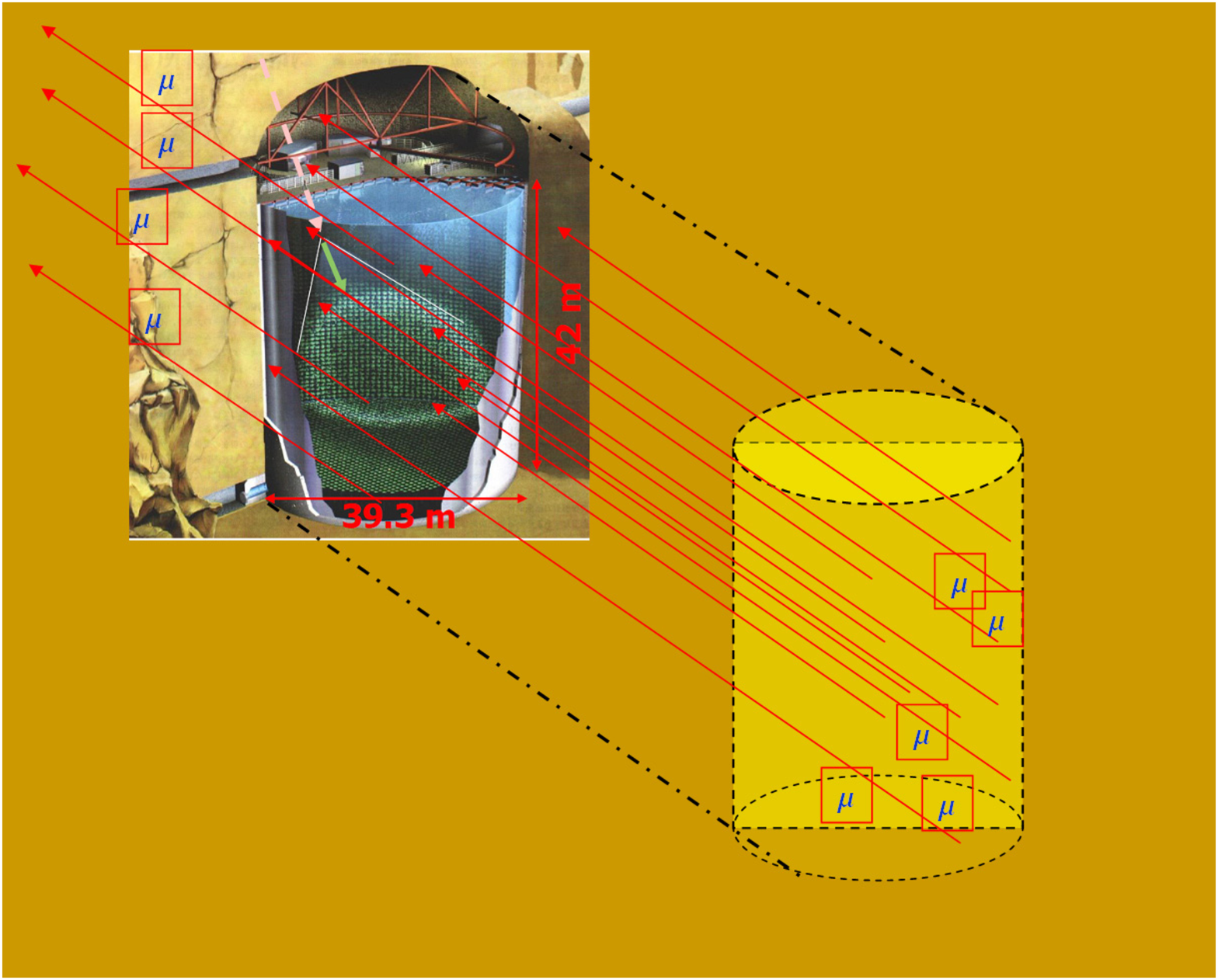}
\includegraphics[angle=0,scale=.17]{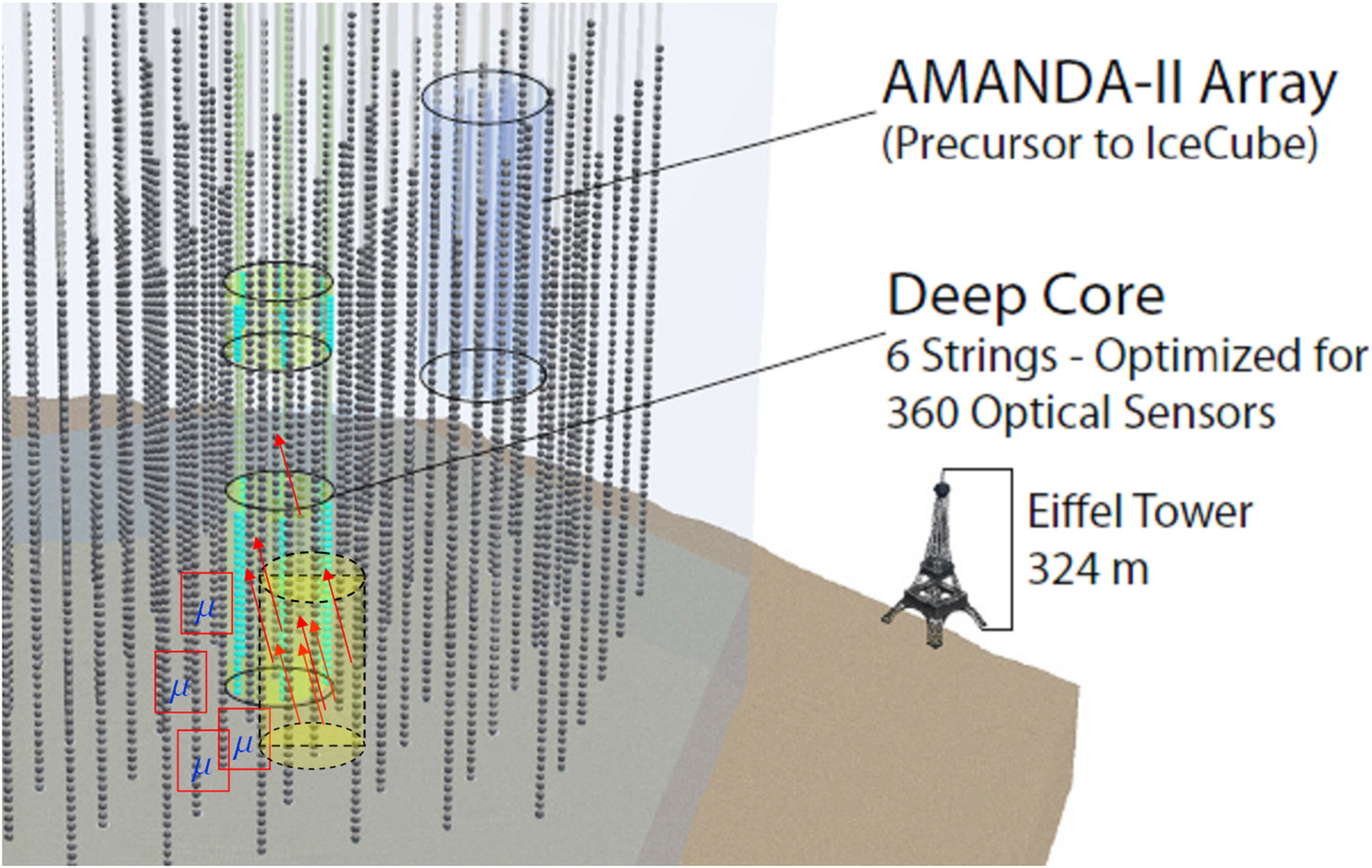}
\caption{ The small size Super-Kamiokande geometry (left figure) for up-going beamed neutrino leading to long (nearly a hundred meter) muons;
 a small fraction of the muons are contained (less that $20$\%). Most (at least $500$\%) are born outside the detector in the nearby rock (rock muons).
On the contrary (right figure) for the larger DeepCore sizes the muon rock (or muon-ice) originated outside the mass detector volume, are just a tiny component (about $25$\%) of the total,
because the larger size of DeepCore respect to a smaller  muon track. }\label{SK-DC}
\end{center}
\end{figure}

\begin{figure}[h]
\begin{center}
\includegraphics[angle=0,scale=.2]{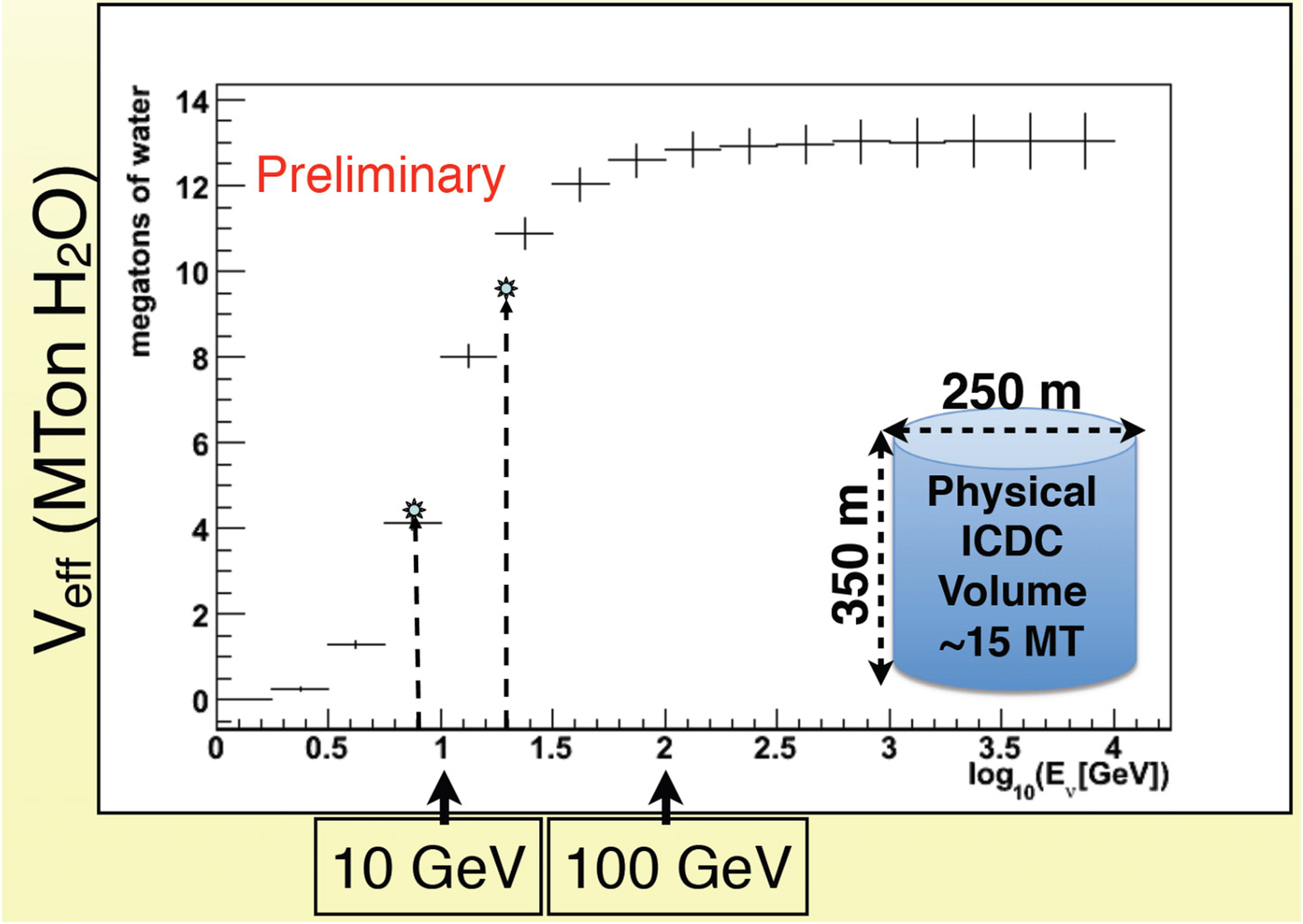}
\includegraphics[angle=0,scale=.212]{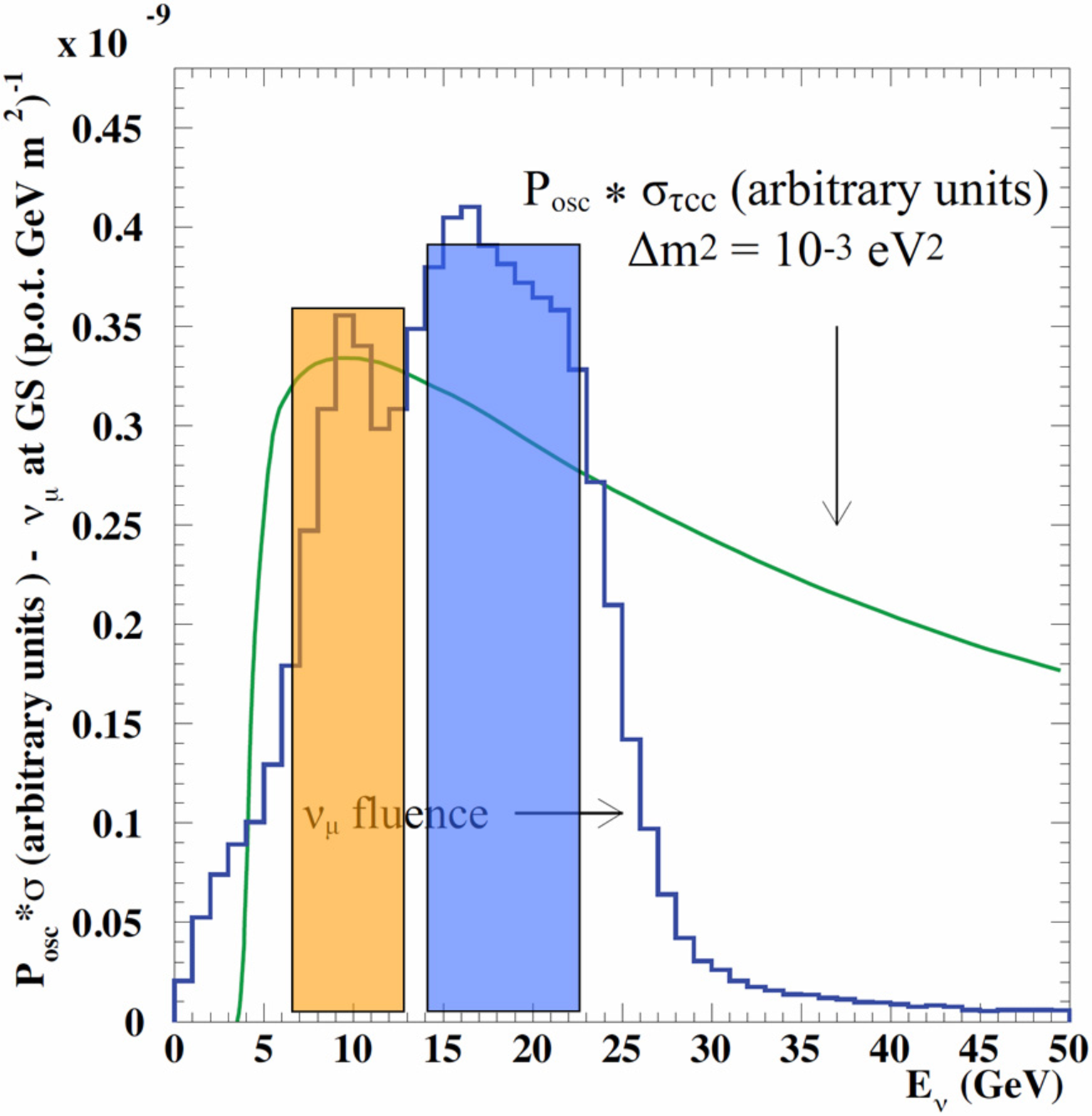}
\caption{ Neutrino Mass detector versus energy for DeepCore. Note that the mass we considered in the extreme range of interest ($20-7.7$ GeV) have been assumed
prudentially at half their shown values in order to avoid any over estimate while we are waiting a more detailed detector calibration.
On the right side the present CERN muon neutrino spectra for OPERA, that we  considered  at two main energy windows (centered at $10$ and $20$ GeV). We do assume the possibility to split these
two bump in quasi-monochromatic  spectra, within $ \frac{\Delta E}{E}$$= \pm 10$\% of their main values, by their spectroscopic bending along the beaming tunnel. Also $ \frac{\Delta E}{E}= \pm 20$\% case has been taken into account. }\label{20}
\end{center}
\end{figure}

\end{appendix}

\clearpage

\end{document}